\documentclass[sn-mathphys,Numbered]{sn-jnl}
\usepackage{graphicx}
\usepackage{multirow}
\usepackage{amsmath,amssymb,amsfonts}
\usepackage{amsthm}
\usepackage{mathrsfs}
\usepackage[title]{appendix}
\usepackage{xcolor}
\usepackage{textcomp}
\usepackage{manyfoot}
\usepackage{booktabs}
\usepackage{algorithm}
\usepackage{algorithmicx}
\usepackage{algpseudocode}
\usepackage{listings}
\usepackage{placeins}

\usepackage{geometry}
\newcommand{\bsx}{\boldsymbol{x}}
\newcommand{\bsX}{\boldsymbol{X}}
\newcommand{\bsu}{\boldsymbol{u}}
\newcommand{\bsn}{\boldsymbol{n}}

\usepackage{graphicx}

\title{Interstitial flow in the chick yolk sac exhibits organ-scale patterns driven by segregated leakage and drainage}
\author[1]{\fnm{Adithya} \sur{Srinivasan}}\email{srini162@purdue.edu}

\author[1]{\fnm{Deshik Reddy} \sur{Putluru}}\email{dputluru@purdue.edu}

\author[2,3]{\fnm{Elizabeth A.V.} \sur{Jones}}\email{liz.jones@kuleuven.be}

\author*[1,4]{\fnm{Hector} \sur{Gomez}}\email{hectorgomez@purdue.edu}

\affil[1]{\orgdiv{School of Mechanical Engineering}, \orgname{Purdue University}, \orgaddress{\city{West Lafayette}, \country{USA}}}
\affil[2]{\orgdiv{Department of Cardiovascular Sciences}, \orgname{KU Leuven}, \orgaddress{ \city{Leuven}, \country{Belgium}}}
\affil[3]{\orgdiv{Department of Cardiology}, \orgname{CARIM School for Cardiovascular Diseases, Maastricht University}, \orgaddress{ \city{Maastricht}, \country{Netherlands}}}
\affil[4]{\orgdiv{Weldon School of Biomedical Engineering}, \orgname{Purdue University}, \orgaddress{ \city{West Lafayette}, \country{USA}}}

\abstract{
Interstitial flow plays a key role in drug delivery, angiogenesis, cancer, and edema. Recent studies have identified connected interstitial pathways that can transport material over long distances. 
However, the spatial scale of physiological interstitial flow remains unclear: despite the existence of connected pathways, flow may be dominated by nearby vascular filtration or extend over longer distances. 
Here, we identify organ-scale interstitial flow patterns in the chick yolk sac and elucidate the mechanisms that determine their spatial scale. 
The yolk sac contains an organ-scale interstitial region that enables large-scale flow patterns to be identified without truncation. 
Our approach combines experimental imaging with computational modeling of coupled blood and interstitial flow in the whole yolk sac. Scaling analysis identifies a hydraulic conductivity ratio that controls the transition from small-scale to organ-scale flow. In organ-scale patterns, we find interstitial flow speed to be substantially greater than the transvascular flow speed. 
Mechanistically, the organ-scale patterns arise from the spatial segregation of the leakage and drainage of interstitial fluid from the vessels. These findings have important implications for biological transport by interstitial flow, suggesting that flow-mediated cues may be sensed locally but generated nonlocally. 
}

\begin{document}

\maketitle

\section{Introduction}
Interstitial flow shapes the transport of solutes, signaling molecules, and drugs in the extravascular space and is therefore important in development, edema, and cancer \cite{desposito_computational_2018, abe_balance_2019, wiig_interstitial_2012, munson_interstitial_2013}. 
Recent evidence suggests that, in addition to the traditionally recognized vascular and lymphatic routes for fluid and solute transport, the interstitium can also act as a long-range
extravascular transport pathway.
Imaging and histological studies have identified macroscopic fluid-filled tissue spaces within the interstitium \cite{benias_structure_2018}, and subsequent studies have reported continuity of these interstitial spaces across tissue and organ boundaries \cite{cenaj_evidence_2021}. 
Further studies have shown body-scale transport of nanoparticles through interstitial pathways \cite{hu_special_2019, liu_interstitial_2022}.  
Together, these studies show that the interstitium can form connected pathways capable of supporting long-distance material transport. 
However, these tracer studies do not show whether physiological interstitial fluid transport actually occurs over long distances, because the observed tracer transport may be driven by mechanisms other than interstitial flow.
The same connected space could support small-scale flow patterns, with short interstitial flow streamlines determined primarily by filtration from nearby vessels, or support large-scale patterns, with long interstitial flow streamlines where the interstitial flow is substantially influenced by filtration from distant vessels.  
Thus, the spatial scale of physiological interstitial flow patterns is unclear.

This spatial scale matters because interstitial flow generates biochemical and biomechanical cues that cells respond to. 
For example, in cancer, interstitial flow guides cell migration in tumors through autologous chemotaxis and other flow-sensitive mechanisms \cite{shields_autologous_2007}. 
This makes the pattern of interstitial flow streamlines important to flow-guided invasion. 
Studies in glioblastoma have shown that the density of tumor-originating interstitial flow streamlines is elevated in regions containing invading cells \cite{carman-esparza_interstitial_2025}.  
Computational models further suggest that interstitial flow generated by leaky tumor vessels and drained by peritumoral lymphatic vessels can create long flow paths that bias tumor cell migration toward lymphatics, which makes flow-guided migration a potential route for metastasis \cite{evje_how_2019}.
Thus, whereas small-scale flow with short streamlines would confine flow-mediated invasion cues near the tumor, large-scale flow with long streamlines could spread soluble factors and directional migration cues away from the tumor, which can potentially facilitate long-distance invasion and metastatic dissemination. 

Earlier attempts to determine the spatial scale of interstitial flow have been limited by the lack of spatially resolved information on interstitial flow patterns. Direct measurements of in vivo interstitial flow are difficult and cannot provide high spatial resolution within complex extravascular geometries \cite{wiig_interstitial_2012}. Resolving the spatial scale of interstitial flow also requires computing flow over a bounded domain whose boundaries are impermeable to interstitial flow, so that complete flow streamlines are contained within the computed domain. 
Otherwise, long streamlines may be truncated by artificial boundaries, making large-scale flow patterns difficult to identify.
However, prior computational studies of in vivo interstitial flow have generally been restricted to tissue sections, where artificial boundaries may make large-scale interstitial flow patterns difficult to identify.

Here, we identify large-scale interstitial flow patterns in the chick yolk sac and elucidate the mechanisms that govern the spatial scale of interstitial flow.
To achieve this, we use experimental imaging combined with computational modeling to resolve blood and interstitial flow across the entire yolk sac. 
Blood flow is modeled as Stokes flow and interstitial flow as Darcy flow, with the two models coupled using Starling's filtration law.  
At the developmental stage studied, the yolk sac is small enough to image and simulate in full and provides an organ-scale bounded interstitial domain. This allows long streamlines to be resolved without truncation by artificial boundaries. 
Thus, any long-range interstitial flow pattern must appear as organ-scale transport across the yolk-sac tissue.  
This setting tests whether the long-distance interstitial transport suggested by prior studies can emerge as  physiological interstitial flow generated by the organ-scale vascular pressure field.

Our analysis uses two complementary length scales, one based on interstitial fluid transport and the other on the amount of vascular information required, to show that physiological interstitial flow in the yolk sac exhibits organ-scale patterns. 
We identify the conductivity ratio, defined as the ratio of interstitial hydraulic conductivity to blood vessel-wall hydraulic conductivity, as the dimensionless parameter controlling these length scales, with increasing values driving the transition from small-scale flow patterns to organ-scale flow patterns.  
In the organ-scale regime, the interstitial flow speeds are substantially larger than transvascular flow speeds.
This result is notable given the common assumption that interstitial flow speeds closely reflect local filtration velocities \cite{kingsmore_mri_2018}.
These organ-scale flow patterns are enabled by the spatial segregation of transvascular leakage and drainage, which causes leaked fluid to travel long distances before reabsorption.

Together, these findings clarify the spatial scale of physiological interstitial flow and provide a concrete mechanism for how this scale is determined.
More broadly, our results suggest that flow-mediated biological cues, including those that guide tumor-cell migration, can be sensed locally but are generated nonlocally by large-scale interstitial flow patterns.
Our findings also suggest that simulations performed on subsections of an entire organ may miss organ-scale patterns and may require specialized boundary conditions to account for interstitial flow in the surrounding tissue \cite{cattaneo_computational_2014, sweeney_modelling_2019, possenti_global_2020}.

\section{Results}
\subsection{Organ-scale flow model of the chick embryo yolk sac} \label{sec: Results sec 1}

\begin{figure}
    \centering
    \includegraphics[width=1\linewidth]{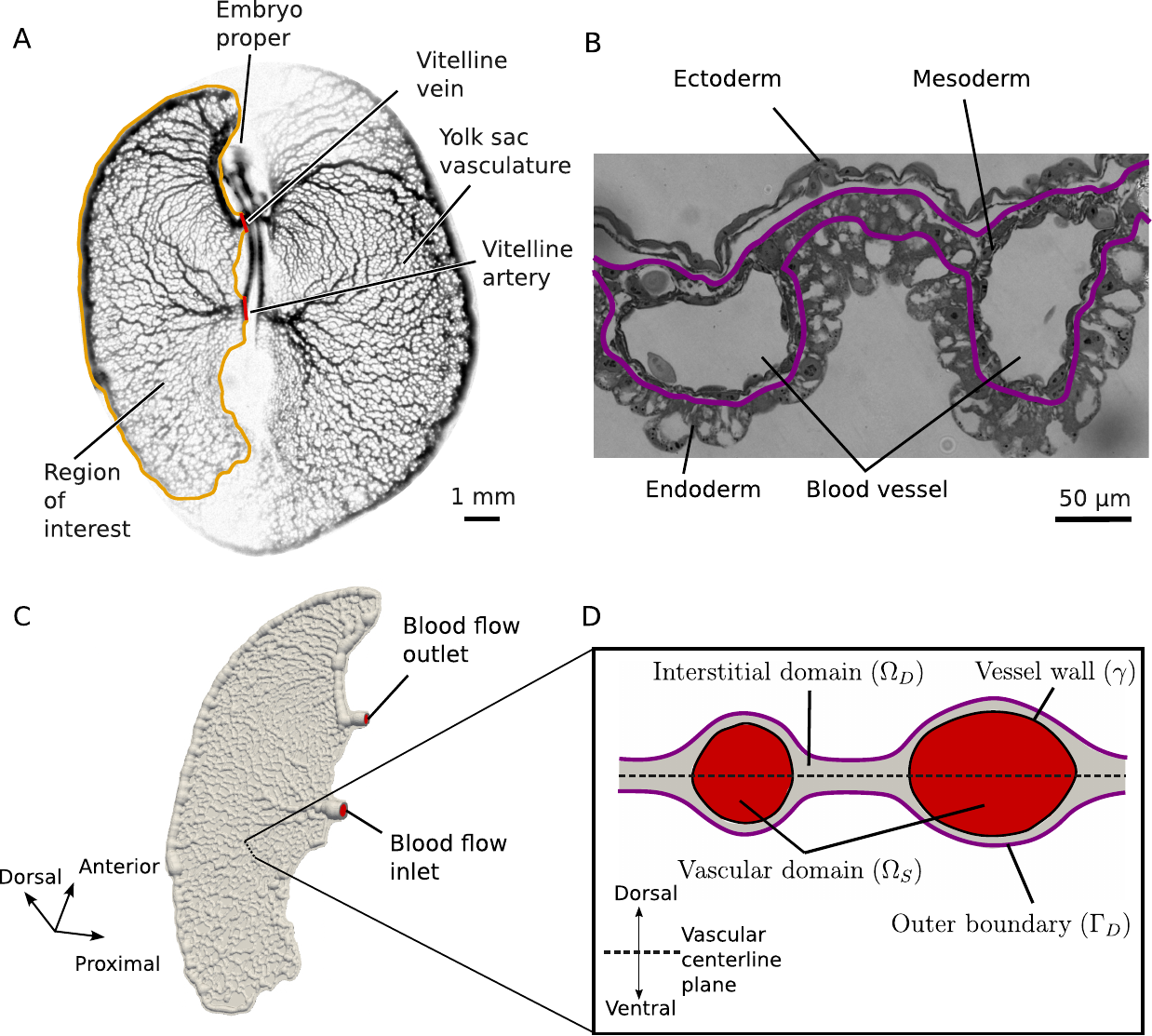}
    \caption{\textbf{Imaging and reconstruction of the chick yolk sac vasculature and interstitium.}
    (A) Top-down microscopy image of the chick embryo yolk sac vasculature. The region of interest (ROI) used for simulations and key anatomical landmarks are indicated. Dark structures indicate blood vessels.
    (B) Representative histological cross-section of the yolk sac tissue, showing the geometry of the interstitial region perpendicular to the vascular centerline plane.
    The section was imaged from a histological slide provided by Dr.~J.~Matthias Starck and prepared during the experiments reported in \cite{starck_morphology_2021}.
    The ectoderm, mesoderm, endoderm, and blood vessels are labeled. The magenta outline marks the boundary of the mesodermal layer. 
    (C) 3D vascular and interstitial geometry with the vascular inlet and outlet marked.
    (D) Cut section of the 3D geometry labeled with the vascular domain $\Omega_S$, the interstitial domain $\Omega_D$, the vessel wall $\gamma$, and the outer interstitial boundary $\Gamma_D$.}
    \label{fig:Overview}
\end{figure}

The yolk sac is an extra-embryonic organ of the chick embryo, and its mesodermal layer contains the developing blood vessel network. 
We refer to the extravascular space within the mesodermal layer as the interstitial region. 
We study blood flow in the vessels and interstitial flow in the surrounding mesodermal tissue.

We first obtain the vascular morphology by injecting chick embryos with dextran to label the perfused vasculature (Methods). 
We image the yolk sac vasculature at  55 hours post-incubation, corresponding approximately to Hamburger--Hamilton (HH) 15 developmental stage.   
Fig.~\ref{fig:Overview}A shows the top-down grayscale image of the vasculature, with key features identified. 
We segment the grayscale image shown in Fig.~\ref{fig:Overview}A to obtain the 2D binary image of the vasculature. In the HH15 developmental stage, the yolk sac vasculature is approximately planar \cite{le_noble_flow_2004}.
We use this planarity to reconstruct a 3D vascular geometry from the 2D binary image using minimal assumptions (Methods).
We embed the 2D vessel geometry into a 3D coordinate system by placing the vessel centerlines in a common plane, which we refer to as the vascular centerline plane.
The constructed 3D geometry preserves the 2D vessel network topology and local vessel diameter, and its orthographic projection onto the vascular centerline plane matches the original 2D image.

Since the interstitial space is difficult to image directly, we construct it from the cross-sectional anatomy of the yolk-sac tissue.
Fig.~\ref{fig:Overview}B shows a representative cross-section of the yolk sac tissue. In the mesoderm, the extravascular region forms a thin tissue layer surrounding the vessels. We therefore model the interstitial region as a perivascular layer of constant thickness around the vasculature, estimated from histological sections \cite{starck_morphology_2021} (Methods).  
Fig.~\ref{fig:Overview}B defines the boundary of the interstitial domain perpendicular to the vascular centerline plane. 
The interstitial domain boundary in the vascular centerline plane is the boundary of the region of interest (ROI) shown in Fig.~\ref{fig:Overview}A; see Methods. The ROI boundary also intersects the vitelline artery and vein, creating inlet and outlet boundaries for the vascular model.

Fig.~\ref{fig:Overview}C shows the reconstructed 3D interstitial and vascular geometry. In this figure, the vascular domain is largely hidden because the interstitial domain forms a continuous tissue volume around it. To visualize the vascular domain more clearly, Fig.~\ref{fig:Overview}D shows a cut section of the 3D geometry, illustrating the vascular domain ($\Omega_S$), the interstitial domain ($\Omega_D$) and their shared interface ($\gamma$). The outer boundary of the interstitial domain ($\Gamma_D$) corresponds to the union of the endoderm-mesoderm boundary, ectoderm-mesoderm boundary, and the ROI boundary in the vascular centerline plane.

We model blood flow as steady, incompressible Stokes flow, yielding the governing equations 
\begin{alignat}{1} \label{eq: Blood flow eqs}
    \nabla\cdot\bsu_S = 0, \quad
    -\nabla p_S + \mu_S\nabla^{2}\bsu_S =\boldsymbol{0} \quad \text{in } \Omega_S.
\end{alignat}
Here, $\bsu_S$ and $p_S$ are the blood velocity and pressure, and $\mu_S$ is the dynamic viscosity of blood. 
We model embryonic chick blood as a Newtonian fluid because its viscosity is approximately constant over physiological shear rates at the HH15 developmental stage \cite{midgett_blood_2015}.
Blood flow is driven by the pressure drop between the vitelline artery and the vitelline vein. 
We neglect cardiac pulsatility because interstitial fluid displacement over one heartbeat is negligible relative to the characteristic length of the yolk sac.
In the interstitial space, we use a porous media flow model based on Darcy's law, which leads to the equations:
\begin{alignat}{1} \label{eq:porous media flow}
    \nabla\cdot\bsu_D = 0, \quad \bsu_D = -\frac{k}{\mu_D}\nabla p_D \quad \text{in } \Omega_D,
\end{alignat}
where $\bsu_D$ and $p_D$ are the interstitial fluid Darcy velocity and pressure, $k$ is the permeability of the interstitial tissue, and $\mu_D$ is the dynamic viscosity of the interstitial fluid. The outer boundary of the interstitial domain ($\Gamma_D$) predominantly coincides with natural anatomical barriers. Therefore, we treat $\Gamma_D$ as impermeable to interstitial flow and impose zero normal interstitial fluid Darcy velocity on it. At the vessel wall $\gamma$, we impose three interface conditions \cite{perktold_mathematical_2009}.  
First, we model the filtration of blood plasma into the interstitial region using Starling's law.
Assuming negligible spatial variation in oncotic pressure across the yolk sac, we omit the oncotic pressure contribution from Starling's law and use the following filtration condition:
\begin{alignat}{1} \label{eq: Starling's law} 
    \bsu_D \cdot \bsn= L_p(p_S - p_D) \quad \text{on } \gamma, \end{alignat} 
where $L_p$ is the hydraulic conductivity of the vessel wall, and $\bsn$ is the unit normal to $\gamma$ pointing into $\Omega_D$. 
Second, we impose the no-slip condition for blood flow at the vessel wall. Finally, the model is completed by imposing mass flux continuity across the interface, i.e., $\bsu_S\cdot\bsn=\bsu_D\cdot\bsn$ on $\gamma$. In our implementation, the mass flux continuity condition is approximated by $\bsu_S\cdot\bsn=0$. We have verified that this approximation produces negligible changes in the results (see Methods) because the interstitial flow is too weak to modify vascular flow. 
This simplification makes the flow model one-way coupled.  
The interstitial pressure does not affect the blood flow solution, but the blood pressure drives interstitial flow through the filtration condition in Eq.~\eqref{eq: Starling's law}.

The estimation of the flow-model parameters is detailed in Methods. 
Throughout this work, we hold the blood-flow viscosity $\mu_S$ and prescribed inlet-outlet pressure drop fixed and vary only the interstitial-flow parameters to determine their impact on the spatial scale of interstitial flow.
These blood flow parameters affect flow magnitudes but not the normalized spatial distribution of vascular and interstitial pressure, and therefore do not alter the spatial scale of interstitial flow.
We first perform a simulation using our best estimates of the interstitial-flow parameters in the yolk sac, which we refer to as the physiological reference case. We then 
vary the interstitial-flow parameters relative to the physiological reference case.

\subsection{Blood flow is pressure-driven and quasi-planar}

\begin{figure}
    \centering
    \includegraphics[width=1\linewidth]{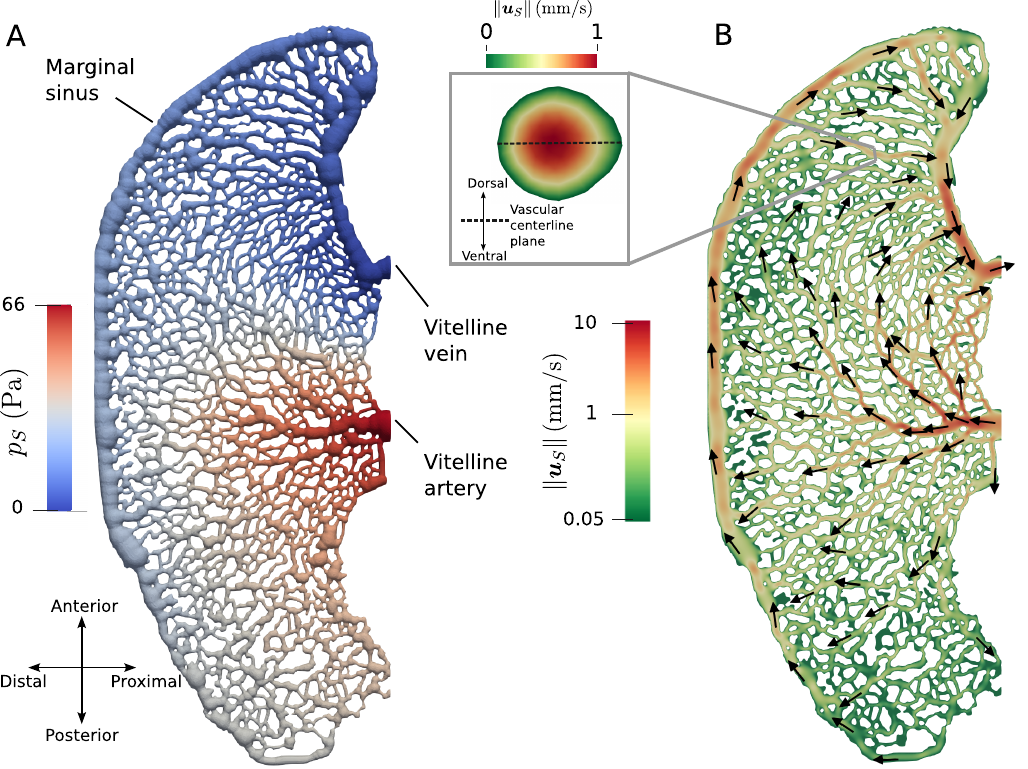}
    \caption{
    \textbf{Blood flow in the chick yolk sac is pressure-driven and quasi-planar.}
    (A) Blood pressure $p_S$ plotted in the 3D vascular geometry. 
    (B) Blood velocity magnitude $\|\bsu_S\|$ plotted on a 2D section of the vascular centerline plane (logarithmic color scale).
    The inset shows the velocity on a vessel section perpendicular to the vascular centerline plane.}
    \label{fig:Blood flow}
\end{figure}

Fig.~\ref{fig:Blood flow}A shows the blood pressure in the entire yolk sac vasculature. Blood pressure is highest at the vitelline artery and decreases along the vascular network to the vitelline vein. This pressure drop drives blood flow. Pressure varies primarily along the vessel centerlines and remains nearly uniform across individual vessel cross-sections (not shown), which suggests that vascular flow is directed primarily along the vascular centerline plane. We study the dimensionality of the flow by measuring the flow dimensionality metric $K_{DV}$, which quantifies the contribution of the dorsal-ventral component of the velocity to the overall velocity field; see Supplementary Section~\ref{app: metrics-defn}. We find $K_{DV}=0.29\%$, which indicates that the blood flow in the yolk sac is quasi-planar.

Fig.~\ref{fig:Blood flow}B shows the blood velocity in the vascular centerline plane, with two dominant flow pathways. 
In the first pathway, blood courses anteriorly from the vitelline artery to the vitelline vein.  
In the second pathway, blood courses through the capillary plexus and feeds into the marginal sinus, which then drains into the vitelline vein.  
These yolk-sac blood flow patterns match those reported in \cite{hogers_intracardiac_1995}. Across straight vessel segments, the blood velocity magnitude is highest near the vessel centerline and decreases toward the vessel wall (inset in Fig.~\ref{fig:Blood flow}B), consistent with pressure-driven laminar flow.

\subsection{Physiological interstitial flow exhibits organ-scale flow patterns}
\label{sec: interstitial flow}
\begin{figure}
    \centering
    \includegraphics[width=\linewidth]{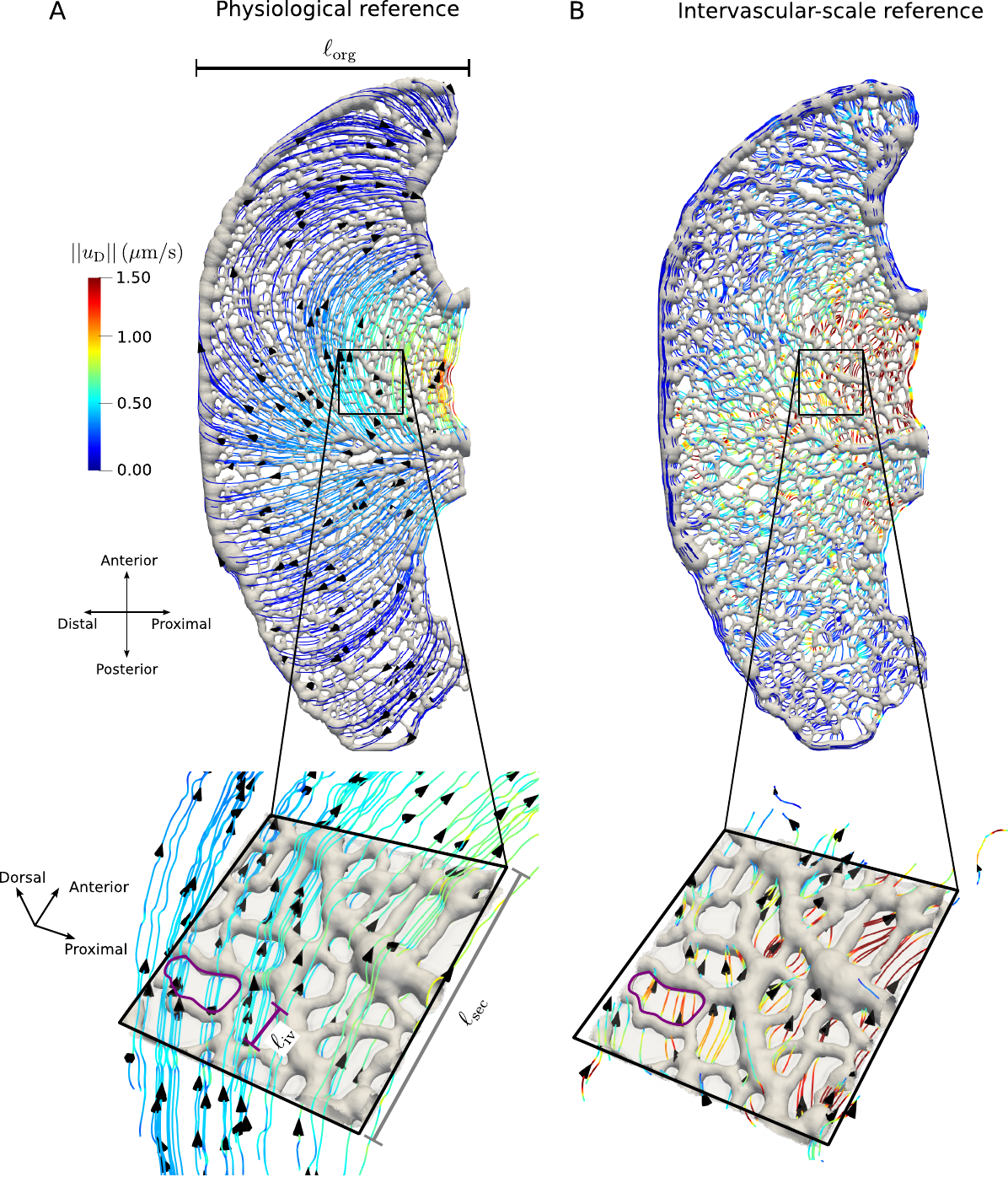}
    \caption{\textbf{Physiological interstitial flow in the chick yolk sac exhibits organ-scale patterns.} Interstitial flow streamlines, colored by the interstitial flow speed $\|\bsu_D\|$, are overlaid on the vascular geometry for the physiological reference simulation (A) and the intervascular-scale reference simulation (B). The yolk-sac extent along the distal-proximal axis is denoted by $\ell_{\rm org}$. The insets show a square organ section with side length $\ell_{\rm sec}$. In the vascular centerline plane, vessels split the interstitial domain into disconnected interstitial regions; see the magenta line and the definition of the characteristic intervascular distance $\ell_{\rm iv}$. The intervascular-scale simulation uses non-physiological parameter values to produce interstitial flow patterns whose length scale is comparable to the intervascular distance $\ell_{\rm iv}$. 
}
    \label{fig:Interstitial flow}
\end{figure}
Fig.~\ref{fig:Interstitial flow}A shows the streamlines of the interstitial flow for the physiological reference simulation overlaid on the vascular network. 
Since interstitial flow is driven by blood pressure, its streamlines broadly follow routes similar to those of blood flow. The streamlines originate from vascular leakage regions near high-pressure vessels around the vitelline artery and drain toward lower-pressure vessels near the vitelline vein and marginal sinus.  
In the vascular centerline plane, vessels split the interstitial domain into disconnected interstitial regions.  
An example interstitial region is shown in the inset of Fig.~\ref{fig:Interstitial flow}A.  
The inset also shows that, in the full 3D geometry, interstitial flow passes above and below
the vessels, connecting interstitial regions that appear disconnected in the vascular centerline plane.
This out-of-plane motion requires a dorsal--ventral velocity component, which manifests as $K_{DV}=9\%$; see Supplementary Section~\ref{app: metrics-defn}. 
This result indicates that physiological interstitial flow exhibits approximately 30-fold greater three-dimensionality than blood flow.
The 3D connectivity gives interstitial flow a continuous pathway across vessel-separated interstitial regions and enables long-distance interstitial fluid transport across the yolk-sac tissue.

Interstitial fluid travels from regions of vascular leakage to regions of interstitial drainage. Those two regions can be differentiated by the sign of the transvascular flux $u_T =\bsu_D\cdot\bsn$. A positive $u_T$ corresponds to leakage into the interstitium, whereas a negative $u_T$ indicates drainage into the vessels. Because flow streamlines connect leakage sites ($u_T>0$) to drainage sites ($u_T<0$), their lengths quantify the spatial extent of interstitial transport.
We therefore define the interstitial transport length, $L_T$, as the mean length of the upper quartile of the streamline-length distribution; see Supplementary Section~\ref{app: metrics-defn}. 
The physiological reference simulation yields $L_T=0.79\,\ell_{\rm org}$, where $\ell_{\rm org}$ denotes the yolk-sac extent along the distal--proximal axis; see Fig.~\ref{fig:Interstitial flow}. This value shows that interstitial transport spans distances comparable to the organ size.

We now introduce another length scale for interstitial flow based on the amount of vascular context required to estimate the organ-scale solution.     
We perform section-scale simulations within subdomains of the yolk sac organ, each constructed  
by extending a square window in the vascular centerline plane across the dorsal--ventral thickness of the tissue, with window side length $\ell_{\rm sec}$ satisfying $\ell_{\rm iv}\leq\ell_{\rm sec}\lesssim\ell_{\rm org}$; see Fig.~\ref{fig:Interstitial flow}. Here, $\ell_{\rm iv}$ is the intervascular distance, which represents the smallest anatomically meaningful scale of flow organization. We define $\ell_{\rm iv}$ as the characteristic separation between neighboring vessel segments, yielding $\ell_{\rm iv}=0.03\,\ell_{\rm org}$; see Supplementary Section~\ref{app: metrics-defn}. The intervascular distance represents the smallest flow scale because the shortest leakage-to-drainage transport paths are set by the separation between nearby vessels.
The premise of the section-scale study is that if a section-scale simulation reproduces the organ-scale solution within the section, then the vascular information inside the section is sufficient to recover the local interstitial velocity and the spatial scale of interstitial flow is no larger than the size $\ell_{\rm sec}$ of the section.  
We perform section-scale simulations for organ sections of different sizes and at different locations. In these simulations, we compute interstitial flow inside a section using only the blood pressure in the vascular network contained within that section; see Supplementary Section~\ref{app: section_scale}.   
At the section boundary, we impose a zero flux condition to exclude the influence of the vascular and interstitial flow information outside the section.
We measure the difference between the section-scale and organ-scale solutions using the normalized error in interstitial fluid velocity, $E_{\rm sec}$.
We compute $E_{\rm sec}$ at multiple locations for each size and measure the mean error across locations.  
We say that a section-scale simulation reproduces the organ-scale solution when the mean $E_{\rm sec}<20\%$.  
For example, for the section size shown in the inset of Fig.~\ref{fig:Interstitial flow}, the mean $E_{\rm sec}$ is 73.5\%, indicating the sections of this size do not contain enough information to reproduce the organ-scale solution.   
As the size of the section increases, the mean $E_{\rm sec}$ decreases because a larger section contains more vascular information to better reproduce the interstitial flow.    
We leverage this trend to define the vascular information length $L_I$, which is equal to the minimum section size $\ell_{\rm sec}$ required to reproduce the organ-scale solution.  
For the physiological reference simulation, we find $L_I =0.91\; \ell_{\rm org}$, indicating that the physiological flow pattern is organ-scale. 
\FloatBarrier
\subsection{Conductivity ratio controls the spatial scale of interstitial flow}
\label{sec : sensitivty analysis}

To better understand the organ-scale patterns in the physiological reference simulation, we next construct the opposite case: an intervascular-scale flow pattern whose length scale is comparable to the intervascular distance $\ell_{\rm iv}$. Such a flow pattern should be reproducible by a section-scale study whose section size $\ell_{\rm sec} \sim \ell_{\rm iv}$.  
For this to occur, the interstitial pressure should closely track the vascular pressure in the organ-scale simulation, because the interstitial pressure predicted by the section-scale simulation is bounded by the narrow range of vascular pressures inside such a small section. 
We quantify this tracking using the vessel-wall averaged absolute transvascular pressure drop $\langle|\Delta p|\rangle$, where $\Delta p = p_S - p_D$; see Supplementary Section~\ref{app: metrics-defn}. As $\langle|\Delta p|\rangle \to 0$, the length scale of flow patterns will decrease.

Scaling analysis indicates that the coupling between interstitial and vascular pressures is controlled by the dimensionless conductivity ratio $R_c=k/(\mu_D L_p t_w)$, which represents the ratio of the interstitial hydraulic conductivity to the vessel-wall hydraulic conductivity. Here, $t_w$ is the vessel-wall thickness.  
Using $R_c$, we rewrite the  filtration law (Eq.~\eqref{eq: Starling's law}) as
\begin{equation} \label{eq: rewritten}
-R_c \nabla p_D \cdot \bsn = 
\frac{p_S - p_D}{t_w} 
\quad \text{on } \gamma, 
\end{equation} 
which shows that decreasing $R_c$ forces the interstitial pressure $p_D$ to closely track the vascular pressure $p_S$.
The prediction of the scaling analysis was confirmed by a high-fidelity simulation with $R_c=10^{-3}R_c^{\rm phy}$, where $R^{\rm phy}_c$ denotes the conductivity ratio in the physiological reference simulation.
In the simulation with reduced conductivity ratio, the average transvascular pressure drop was $\langle |\Delta p| \rangle = 0.05~\mathrm{Pa}$ compared to $\langle| \Delta p|\rangle = 1.69~\mathrm{Pa}$ in the physiological reference simulation. 
The resulting vascular information length was $L_I=1.5\,\ell_{\rm iv}$, indicating that the interstitial flow at a given location can be recovered using vascular information spanning only a few intervascular distances.
Hence, we refer to this case as the intervascular-scale reference.

Fig.~\ref{fig:Interstitial flow}B shows the interstitial flow streamlines for the intervascular-scale reference, which are in sharp contrast to those of the physiological reference simulation.  
The intervascular-scale reference streamlines (inset of Fig.~\ref{fig:Interstitial flow}B) typically originate from one vessel and end at the closest vessel.
The streamlines remain confined to individual interstitial regions, rarely passing dorsally or ventrally to the vessels.   
Consequently, the flow is quasi-planar, with the flow dimensionality metric $K_{DV}=2\%$.   
Since the streamlines remain confined to individual interstitial regions, they span only a few intervascular distances, resulting in a short interstitial transport length $L_T = 2.1 \;\ell_{\rm iv}$. 
This short transport length explains why the vascular information over a similarly short range ($L_I=1.5\,\ell_{\rm iv}$) is sufficient to determine the local interstitial flow.

\begin{figure}
    \centering
    \includegraphics[width=1\linewidth]{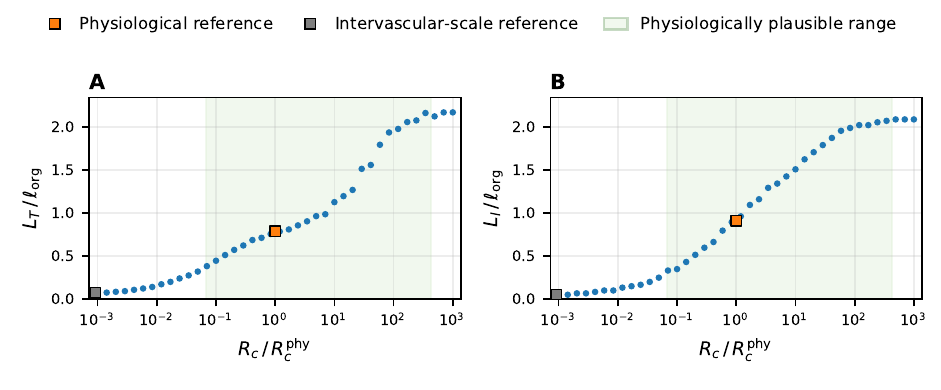}
    \caption{\textbf{The conductivity ratio controls the spatial scale of interstitial flow.}
Variation of (A) the interstitial transport length $L_T$ and (B) the vascular information length $L_I$ with the conductivity ratio $R_c$, for 40  cases over the range $[0.001R^{\rm phy}_c, 1000 R^{\rm phy}_c]$, where  $R^{\rm phy}_c$ is the value of $R_c$ in the physiological reference simulation.  
The green shaded region represents the physiologically plausible range of interstitial flow parameter values.} 
    \label{fig:sensitivity}
\end{figure}

The contrast between the physiological reference simulation and the intervascular-scale reference shows that decreasing the conductivity ratio $R_c$ shifts interstitial flow from organ-scale to intervascular-scale organization.
We next test whether $R_c$ controls this transition continuously by varying $R_c$ over $[10^{-3}R_c^{\rm phy},10^3R_c^{\rm phy}]$ and computing the interstitial transport length $L_T$ and vascular information length $L_I$; see Supplementary Section~\ref{app: sensitivty_analysis}.
For a fixed blood flow solution, scaling analysis shows that the interstitial pressure $p_D$, $L_I$, and $L_T$ depend on the interstitial flow parameters $(L_p,k,\mu_D)$ only through $R_c$.
Thus, sweeping $R_c$ captures the variation of all the interstitial-flow parameters.
The sweep also includes the physiologically plausible range of flow parameters, determined from the estimated uncertainty in the underlying interstitial-flow parameters of the chick yolk sac.
Fig.~\ref{fig:sensitivity} shows that $L_T$ and $L_I$ increase with $R_c$, indicating that increasing $R_c$ drives a continuous transition from intervascular-scale to organ-scale flow organization.
Across the physiologically plausible range of interstitial flow parameter values, both lengths $L_I$ and $L_T$ are much larger than the intervascular distance $\ell_{\rm iv}$ and are comparable to the organ-scale length $\ell_{\rm org}$.
Thus, interstitial flow remains organ-scale across the physiologically plausible parameter range.
Supplementary Section~\ref{app: Rc_metrics} shows that the flow dimensionality metric $K_{DV}$ and the transvascular pressure drop $\langle|\Delta p|\rangle$ also increase monotonically with $R_c$.
This finding shows that these metrics can be understood as signatures of the spatial scale of flow.

\subsection{Interstitial flow speed substantially exceeds transvascular velocity in organ-scale flow patterns }
In intervascular-scale flow patterns, interstitial fluid leaks from one vessel and drains into a nearby vessel.  
Therefore, interstitial fluid speed should be determined primarily by local transvascular exchange and be comparable to the transvascular velocity magnitude. 
We quantify this comparison using the dimensionless number $R_v$, defined as the ratio of vessel-wall averaged interstitial fluid speed to vessel-wall averaged transvascular velocity magnitude; see Supplementary Section~\ref{app: metrics-defn}.   
The intervascular-scale reference yields $R_v=2.3$, indicating that interstitial and transvascular flow speeds are of the same order.
In contrast, the physiological reference simulation yields $R_v=55$, indicating that interstitial flow is much faster than transvascular flow.  
Consistent with this trend, Supplementary Section~\ref{app: Rc_metrics}  shows that $R_v$ increases monotonically with the conductivity ratio $R_c$, supporting the interpretation of $R_v$ as a signature of organ-scale flow patterns.

\subsection{Organ-scale interstitial flow patterns are enabled by segregation of transvascular leakage and drainage}

\begin{figure}
    \centering
    \includegraphics[width=1\linewidth]{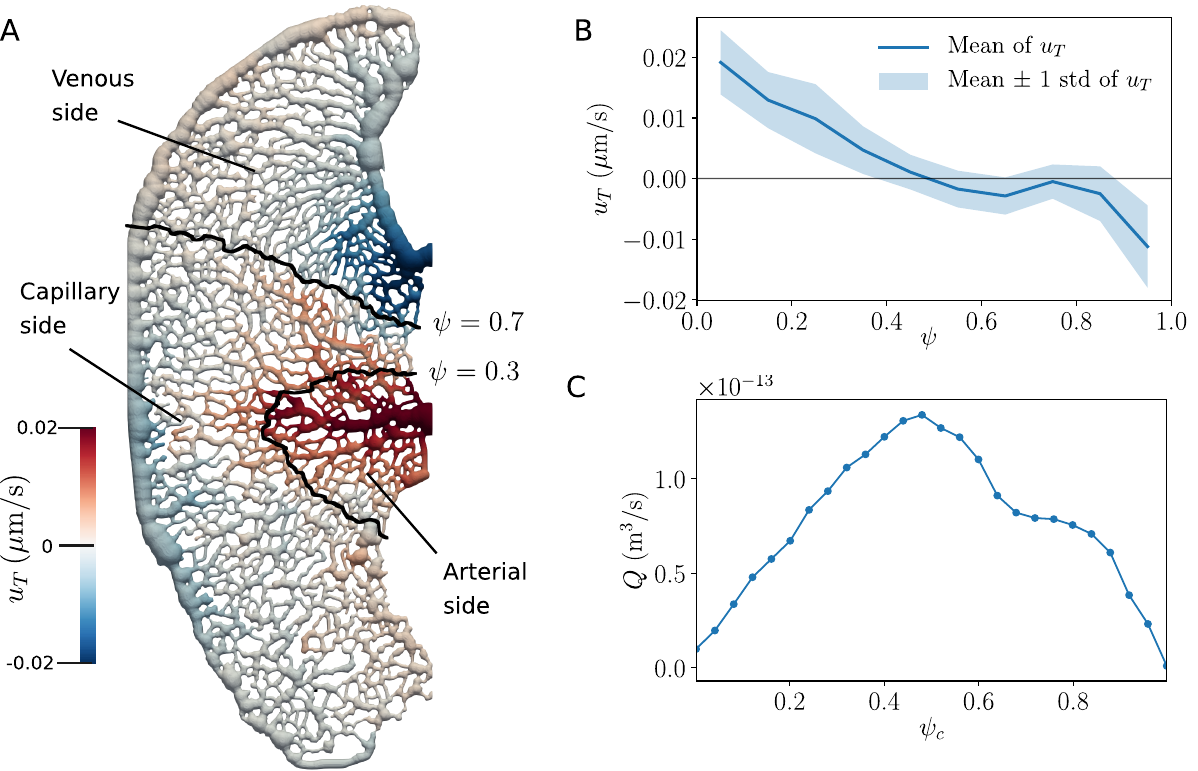}
    \caption{\textbf{Segregation of transvascular leakage and drainage drives interstitial flux accumulation.} (A) Transvascular velocity $u_T$ plotted on the vessel wall $\gamma$ for the physiological reference simulation. The arterial-side, capillary-side, and venous-side interstitial tissue regions are marked using the isosurfaces of the arterio-venous coordinate $\psi$. 
    (B) Plot of $u_T$ against $\psi$ on the surface mesh nodes of $\gamma$. Surface mesh nodes on $\gamma$ are grouped into 10 bins based on their $\psi$ values. The solid line indicates the mean of $u_T$ in each bin, and the shaded region shows the $\pm 1$ standard deviation of $u_T$ within each bin. 
    (C) Plot of interstitial fluid flux $Q$ passing through isosurfaces of $\psi$, given by $\psi = \psi_c$. 
    }
    \label{fig:Transvascular flow}
\end{figure}

We investigate the mechanism underlying organ-scale flow patterns by asking why interstitial flow streamlines are long and how interstitial flow can be much faster than transvascular flow. 
Fig.~\ref{fig:Transvascular flow}A shows $u_T$ on the vessel walls for the physiological reference simulation.    
Leakage ($u_T>0$) is concentrated near the vitelline artery, whereas drainage ($u_T<0$) is concentrated near the vitelline vein.  
To quantify this spatial trend, we define an arterio-venous coordinate $\psi$ in the interstitial space; see Supplementary Section~\ref{app: metrics-defn}.
We compute $\psi$ by smoothly extending the blood pressure $p_S$ from the vessel wall into the interstitial domain.
Fig.~\ref{fig:Transvascular flow}A shows that $\psi$ varies continuously from $0$ near the vitelline artery to $1$ near the vitelline vein and partitions the interstitial domain into the arterial-side region ($\psi<0.3$), the capillary-side region ($0.3\leq\psi\leq0.7$), and the venous-side region ($\psi>0.7$). 
Fig.~\ref{fig:Transvascular flow}B  plots the transvascular velocity $u_T$ along the arterio-venous coordinate $\psi$.   
As $\psi$ increases from the arterial side to the venous side, the average $u_T$ decreases and transitions from leakage to drainage.   
Thus, leakage and drainage regions are spatially segregated along the arterio-venous axis.  
Since streamlines connect leakage and drainage sites, this segregation lengthens their transport paths and increases the vascular context required to determine the flow, producing organ-scale values of $L_T$ and $L_I$.  
Consistent with this mechanism, Supplementary Section \ref{app: Rc_metrics} shows that the strength of spatial segregation increases with the conductivity ratio $R_c$.

Since the leakage and drainage are segregated in the physiological reference simulation, interstitial fluid that enters near the arterial side must be transported across the tissue before it exits near the venous side. 
Consequently, the flux through a cross-section in the capillary-side region reflects fluid accumulated from upstream arterial-side leakage before it is removed downstream by the venous-side drainage. 
To quantify this accumulation of interstitial fluid flux, we compute the interstitial fluid flux $Q$ passing through the isosurfaces of the arterio-venous coordinate $\psi$, which serve as cross-sections ordered along the arterio-venous axis; see Supplementary Section~\ref{app: metrics-defn}.   
Fig.~\ref{fig:Transvascular flow}C shows that the interstitial fluid flux $Q$ increases in the arterial-side region, peaks in the capillary-side region, and decreases back toward zero in the venous-side region.
This trend demonstrates that the spatial segregation of leakage and drainage drives the accumulation of interstitial flux along the arterio-venous axis.       
This accumulated flux reflects the cumulative contribution of transvascular filtration from multiple vessels. 
To relate flux accumulation to interstitial fluid speed without mixing the two dominant flow pathways, we analyze the artery-to-vein pathway in a restricted tissue subregion; see Supplementary Section~\ref{app: flux accumulation}. 
Along this pathway, the average interstitial fluid speed at the capillary-side region is $11\times$ larger than the interstitial fluid speed near the vitelline artery, where it is comparable to the local transvascular leakage.  
This increase in interstitial fluid speed follows the same trend as the interstitial flux $Q$. 
This correspondence suggests that accumulation of interstitial fluid flux amplifies interstitial fluid speed and enables it to substantially exceed transvascular velocity in organ-scale flow patterns, as reflected by the high value of $R_v$. 
Further, Supplementary Section \ref{app: Rc_metrics} quantifies the strength of accumulation of interstitial flux and shows that it increases with the conductivity ratio $R_c$, indicating that stronger accumulation accompanies the transition toward organ-scale flow patterns. 
Together, these results show that the segregation of leakage and drainage produces organ-scale flow organization through two related effects: it lengthens interstitial transport paths and accumulates interstitial flux from multiple vessels, thereby increasing $L_T$ and $R_v$.

\section{Discussion}
Experimental studies have reported long-distance transport through connected interstitial pathways \cite{liu_interstitial_2022,cenaj_evidence_2021}, raising the question of whether physiological interstitial flow is organized over similarly large scales.
Our results show that a connected organ-scale interstitial region can support either intervascular-scale or organ-scale interstitial flow patterns, with the ratio of interstitial to vessel-wall conductivity controlling the transition between these regimes.  
Under physiological conditions, interstitial flow in the chick yolk sac exhibits organ-scale flow patterns. 
These organ-scale patterns form because transvascular leakage and drainage are spatially segregated along the arterio-venous axis.

More broadly, if similar large-scale flow patterns persist in other tissues, our results suggest that flow-sensitive biological cues may be local in how they are sensed, but are nonlocal in how they are generated. 
One such cue is the interstitial fluid speed itself, which plays a role in angiogenesis \cite{abe_balance_2019,srinivasan_computational_2024} and cell migration \cite{shields_autologous_2007, paspunurwar_decoding_2024} through mechanotransduction. 
In organ-scale patterns, we show that interstitial fluid speed is not determined by the local transvascular flow from nearby vessels. 
Instead, the fluid leaked from multiple upstream regions accumulates to produce interstitial fluid speed much larger than transvascular flow. 
This result is important because leakage patterns are sometimes used as supporting evidence for reconstructed interstitial fluid speed fields, implicitly assuming that local leakage and local interstitial fluid speed are closely related \cite{kingsmore_mri_2018}.  
Our results suggest that this assumption should be applied with caution when interstitial flow patterns are large-scale.  
A second such cue is the exposure to soluble factors that are transported by interstitial flow. 
In organ-scale flow patterns, the exposure to these soluble factors may depend on the pathlines that deliver fluid and solutes from distant upstream locations. 
This mechanism could help interpret recent studies of cell invasion in glioblastoma. In \cite{carman-esparza_interstitial_2025}, they found that the tumor-originating pathline-density metric, defined as the density of tumor-derived interstitial-flow pathlines passing through each local region, and the local interstitial flow speed were elevated in regions containing invading tumor cells.   
Our results suggest that regions with high tumor-originating pathline density may mark downstream regions where tumor derived fluid from many upstream regions accumulates.  
Such accumulation could increase local interstitial fluid speed and exposure to tumor-derived soluble factors.  
Thus, tumor-originating pathline density may identify regions where the migratory cues become concentrated due to the spatial organization of interstitial flow sources and sinks.

Our findings also have important implications for how interstitial flow is computed in vivo. 
Our analysis defines the vascular information length as the minimum spatial extent of vascular context required to reproduce the organ-scale interstitial flow. 
Our results show that simulations performed with tissue sections smaller than this length exclude vascular information needed to determine the local interstitial flow and therefore fail to reproduce the organ-scale solution accurately.  
In this work, the relatively small size of the yolk sac enabled us to completely capture the organ-scale interstitial flow patterns. 
In larger tissues with more complex vascular geometries, however, full-organ imaging and computation are often not feasible and interstitial flow simulations are typically performed on smaller sections with artificial outer boundaries \cite{cattaneo_computational_2014,possenti_computational_2019, desposito_computational_2018,sweeney_modelling_2019}.  
Our results suggest that if these tissues exhibit organ-scale flow organization similar to the yolk sac, such section-scale simulations  require boundary conditions that represent the influence of vascular filtration outside the simulated domain.    

The organ-scale flow patterns identified in the yolk sac suggest that large-scale interstitial flow can also occur in other tissues.  
Whether these large-scale patterns generalize to other tissues depends on how tissue-specific features modify the spatial segregation of interstitial flow sources and sinks observed in the yolk sac. 
One important difference between the chick yolk sac and many mature tissues is the presence of lymphatic drainage.  
The absence of lymphatics in the yolk sac was advantageous for identifying the mechanism of interstitial flow because transvascular leakage and drainage were the only sources and sinks of interstitial flow.  
This removed the uncertainty associated with the locations and hydraulic properties of a separate lymphatic drainage network and allowed us to implicate the segregation of vascular leakage and drainage as the driver of organ-scale flow patterns.  
However, this simplified source--sink structure limits direct generalization to tissues with lymphatics, which act as an additional, spatially distributed interstitial flow sink.  
How lymphatic drainage affects the spatial scale of interstitial flow remains unclear.  
Dense and spatially distributed lymphatic networks could shorten interstitial transport paths and decrease the spatial scale of flow, whereas spatially concentrated lymphatic drainage could preserve or even generate large-scale patterns.  
Future work could therefore incorporate organ-scale lymphatic networks into models of interstitial flow to determine how the topology of the lymphatic network interacts with transvascular leakage to control the spatial scale of flow.  
Another important difference between the yolk sac and other mature tissues is that the yolk sac has a small and approximately planar vasculature, which makes organ-scale imaging and computation tractable.  
To generalize our findings to other mature tissues, future work could explore whether these large-scale patterns persist in larger tissues with dense, three-dimensional vascular networks.  
In such tissues, the complex arrangement of arteries, capillaries, and veins may alter the spatial organization of leakage and drainage, which may affect the scale of interstitial flow. 
Studying this question requires organ-scale vascular images of the large tissue and computational methods capable of resolving blood and interstitial flow at that scale.  
Recent advances in three-dimensional microvascular imaging \cite{todorov_machine_2020} and hybrid discrete--continuum models of blood pressure can make such computations tractable  \cite{sweeney_threedimensional_2024}.           
The chick chorioallantoic membrane can provide an accessible in vivo system to test the generalizability of the results presented in this work because it contains a larger and denser vascular network as well as lymphatic drainage \cite{richard_direct_2018,nowak-sliwinska_chicken_2014}. Applying the present framework to this system would enable us to study the combined effects of vascular complexity and lymphatic drainage on the spatial scale of interstitial flow.

In summary, we have elucidated the mechanisms that determine the spatial scale of interstitial flow.  
In particular, our analysis identifies the ratio of interstitial conductivity to vessel-wall hydraulic conductivity as the key parameter that controls the spatial scale of interstitial flow.  
Since both conductivities can be modified through interventions that target the extracellular matrix or vessel wall, this mechanism could be used as a potential biophysical lever to design therapeutic strategies to improve drug delivery in tumor microenvironments by promoting long-range interstitial fluid transport.

\section{Methods}
\subsection{Definition of region of interest} \label{sec: ROI}
\begin{figure}
    \centering
    \includegraphics[width=\linewidth]{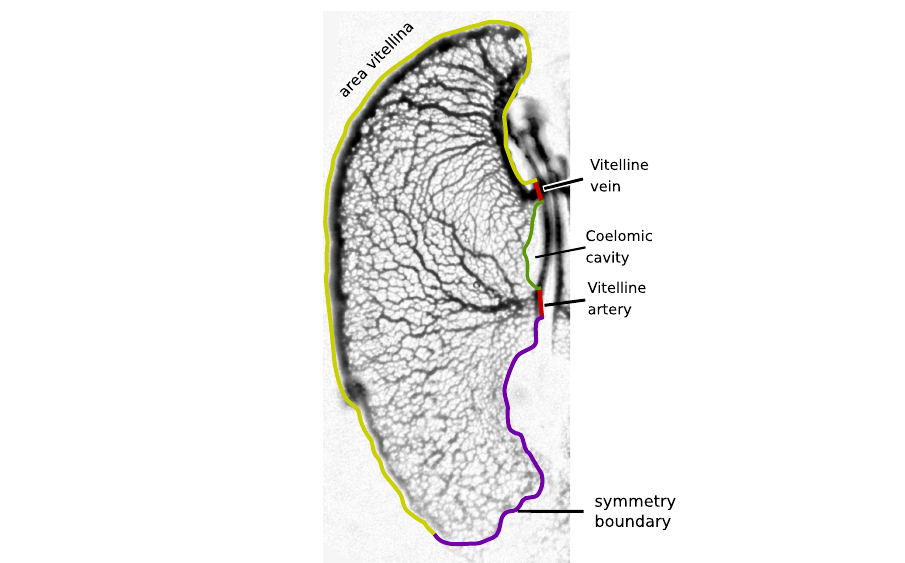}
    \caption{\textbf{Definition of the region of interest in the yolk sac.} Top-down microscopy image of the HH15 yolk-sac vasculature marked with the ROI boundary in the vascular centerline plane. The boundary segments are colored by their anatomical interpretation. The yellow segment is the area vitellina boundary, the green segment is the coelomic cavity boundary, and the purple segment is the approximate symmetry boundary between the left and right halves of the embryo.}
    \label{fig:ROI}
\end{figure}

The region of interest (ROI) defines the boundary of the interstitial domain on the vascular centerline plane and the vascular inlets and outlets. 
We chose the ROI to allow physiologically grounded boundary conditions for interstitial flow.
In the vascular centerline plane, the ROI boundary is divided into three segments (Fig.~\ref{fig:ROI}). 
The yellow and green segments border the area vitellina and the coelomic cavity, respectively. These anatomical interfaces provide natural impermeable boundaries for interstitial flow. The purple segment below the vitelline artery is not a natural boundary but represents the separation between the two halves of the embryo. The vessels outside the purple boundary are not stained, indicating a lack of perfusion in this region. We therefore assume that blood flow through vessels cut by this segment is negligible. Because the vasculature below the vitelline artery is not visible, reconstructing the interstitial geometry and modeling interstitial flow in this region would be difficult.
In principle, some interstitial flow could occur across the purple segment.  
However, because the yolk-sac geometry is approximately symmetric between the left and right halves of the embryo, we assume the net interstitial fluid flux across this segment is negligible. 
Thus, we label this segment as a symmetry boundary and restrict our ROI to a single half of the yolk sac.

\subsection{Imaging of the vascular network}

Fertilized chicken eggs (\textit{Gallus gallus}) were incubated at $38.5\,^\circ\mathrm{C}$ and $\sim 60\%$ humidity on their sides for 55 hours. Eggs were cracked into plastic weigh dishes. Embryos were then humidified with PBS and returned to a $37\,^\circ\mathrm{C}$ incubator for 30 minutes to ensure strong stage-appropriate blood flow was present. Embryos were then injected into the sinus terminalis with fluorescent-tagged dextran solutions (Dextran, Oregon Green\texttrademark{} 488; $70\,\mathrm{kDa}$, anionic; Thermo Fisher Scientific, D7172) using pulled quartz needles attached to a Picospritzer III micro-injector. For all injections, a small amount of blue food dye was added to each dye mixture to enable us to detect successful intravascular injection. After injection, embryos were rehydrated with PBS, then returned to the incubator for 5 to 10 minutes. The embryos were then imaged on a stereomicroscope (Zeiss Discovery V12, Axiocam MRc) equipped with a fluorescent lamp (HXP 120c) at $8\times$ magnification. Several overlapping images were taken and then stitched together to image the entire yolk sac vasculature (Fig.~\ref{fig:Overview}A). An image of a stage micrometer was also taken at the same settings to allow for proper scaling. We estimate the embryonic stage based on the maturation of the heart and the vascular plexus.
We use FIJI \cite{schindelin_fiji_2012} to segment the grayscale yolk sac vasculature image into a 2D binary vessel mask. 
We first crop the image to the ROI, enhance contrast, and apply background subtraction. 
We then perform automatic local thresholding \cite{phansalkar_adaptive_2011} to obtain an initial binary image.  
To remove segmentation artifacts, we use a median filter to eliminate mislabeled isolated pixels.
We then perform a morphological closing with a disk structuring element to close small gaps in vessels. 
Finally, we manually annotate thin capillaries missed by the automatic pipeline. 
We visually inspect the mask overlay on the grayscale image to confirm that vessel boundaries were captured. 
We verify that the vessel network is fully connected by confirming that the final mask comprises a single connected component.

\subsection{Reconstruction of 3D vascular and interstitial geometry}
To construct the 3D vascular geometry, we build a 3D binary lumen from the 2D binary mask.  
This construction preserves the observed 2D network topology and a spatially varying local vessel caliber under minimal geometric assumptions. 
First, we use the FIJI Local Thickness workflow on the 2D binary mask  \cite{hildebrand_new_1997}. The local thickness field 
$\tau(X,Y)$ is defined as the diameter of the maximally
inscribed disk corresponding to pixel $(X,Y)$, which is the largest disk fully contained within the vessel mask that contains the pixel $(X,Y)$. Next, we compute the distance ridge of the binary mask, which is the set of pixels where the Euclidean distance-to-boundary field attains a local maximum. 
The pixels in the distance ridge also correspond to the centers of maximally inscribed disks. 
The union of disks centered on the distance ridge with diameters $\tau(X,Y)$ reconstructs the 2D mask up to pixelization. We reconstruct the 3D binary vascular mask $\hat{\Omega}_S$ by taking the union of spheres centered at distance ridge pixels with diameter $\tau(X,Y)$: 
\begin{equation}
    \hat{\Omega}_S
    =
    \bigcup_{(X,Y)\in \mathrm{DR}}
    B\!\left(\,(X,Y,0),\,\frac{1}{2}\tau(X,Y)\right),
\end{equation}
where $B(c,r)$ denotes a sphere with center $c$ and radius $r$ and DR denotes the distance ridge. We leverage the approximate planarity of the yolk sac vascular network to place the spheres on the vascular centerline plane ($Z=0$). 
Since the equatorial disk of each sphere in the $Z=0$ plane matches the corresponding maximally inscribed disk in the 2D binary mask, the orthographic projection of $\hat{\Omega}_S$ onto the vascular centerline plane recovers the original 2D binary image.
This reconstruction method is similar to previous methods that have been employed to reconstruct the 3D vessel geometry from planar confocal images of the retinal vasculature \cite{mirzapour-shafiyi_numerical_2021}.

To create the 3D geometries of the vascular and interstitial domains, we first build surface mesh representations of the vessel wall $\gamma$ and outer interstitial boundary $\Gamma_D$ from the 3D binary vascular mask $\hat{\Omega}_S$.  
Because this construction uses both voxel indices and physical coordinates, we denote voxel indices on the Cartesian grid by $\bsX=(X,Y,Z)$ and physical coordinates by $\bsx=(x,y,z)=\delta\bsX$, where $\delta$ is the voxel spacing.  
We represent the domain boundaries using signed distance functions (SDFs).  
For a surface $\Sigma$ enclosing a domain $\omega$, the SDF $\phi(\bsX)$ is the signed distance to $\Sigma$, expressed in physical units. We define $\phi<0$ inside $\omega$, $\phi=0$ on $\Sigma$, and $\phi>0$ outside $\omega$.

First, we compute the SDF of the vessel wall, denoted by $\phi_\gamma$, from the binary vascular mask $\hat{\Omega}_S$ using the Euclidean distance transform in SciPy \cite{virtanen_scipy_2020}.  
We apply Gaussian smoothing to $\phi_{\gamma}$ to smooth sharp voxel ridges. 
Next, we use $\phi_{\gamma}$ to construct the outer interstitial boundary $\Gamma_D$.  
The representative histological section (Fig.~\ref{fig:Overview}B) shows that the interstitial region forms a continuous layer that surrounds the vessels and retains a comparable thickness in the avascular regions between them. 
We therefore model the interstitial domain $\Omega_D$ as a continuous layer with constant perivascular thickness $t$ surrounding the vasculature (see Fig.~\ref{fig:Overview}D).  
To estimate $t$, we image histological sections of avian embryo yolk sacs prepared as part of the experiments reported in \cite{starck_morphology_2021}, and measure the perivascular thickness in 5 embryos at 8 locations per embryo ($n=40$).  
At each location, we define the thickness as the distance from the vessel wall to either the mesoderm--ectoderm interface or the mesoderm--endoderm interface.  
Across all measurements, the mean thickness is $17.89\,\mu$m, with a standard deviation of $5.66 \,\mu$m. We set $t$ equal to the measured mean.  
To construct this constant-thickness interstitial layer, we treat vascular and avascular regions within the ROI separately. 
In vascular regions of the ROI, $\Gamma_D$ lies at an offset distance $t$ from the vessel wall $\gamma$ and is identified as the level set $\phi_{\gamma} = t$.  
In avascular regions of the ROI, we construct a continuous interstitial layer using a slab of thickness $2t$ centered on the vascular centerline.  
The boundary of this slab is captured by the SDF $\phi_s$, obtained by extruding the ROI footprint to a total thickness $2t$ along the dorsal--ventral axis. 
We construct the SDF of the outer interstitial boundary, denoted by $\phi_D$, by combining the vascular offset and the slab:  

\begin{equation}   
\phi_D(\bsX) = \text{min}(\phi_{\gamma}(\bsX) - t, \phi_s(\bsX)).  
\end{equation}  
We extract triangular surface meshes for $\gamma$ and $\Gamma_D$ from the level sets $\phi_{\gamma}=0$ and $\phi_D=0$, respectively,  using the marching cubes algorithm \cite{walt_scikit-image_2014}.  
We then apply Taubin smoothing to obtain smooth triangular meshes.     

Finally, we identify the vascular inlet and outlet surfaces, $\Gamma_i$ and $\Gamma_o$, as the ROI cuts through the vitelline artery and vein (Fig.~\ref{fig:Overview}A) and represent them as planar cross-sections with unit normals parallel to the anterior-posterior axis.  
We use the identified surfaces ($\gamma$, $\Gamma_D$, $\Gamma_i$ and $\Gamma_o$)  to generate volumetric meshes for the vascular and interstitial domains.  
To verify that smoothing does not distort the original geometry excessively, we confirm that the relative volume difference between $\Omega_S$ and $\hat{\Omega}_S$ is 3.5\%.  

\subsection{Justification of the flow model}

Here, we detail the assumptions that we used in the flow models presented in Sec.~\ref{sec: Results sec 1}. To model blood flow, we use the incompressible Stokes equations [Eq.~\eqref{eq: Blood flow eqs}] instead of the Navier--Stokes equations because the Reynolds number is small. 
We model blood as a Newtonian fluid with a constant viscosity, following previous work showing that embryonic chick blood viscosity is approximately constant over physiological shear rates \cite{midgett_blood_2015}. 
Although the blood flow in the yolk sac oscillates with the heartbeat, we use a steady blood flow model because our primary goal is to compute the blood pressure field that drives interstitial flow. Because the heartbeat timescale is $\sim 1\,\mathrm{s}$, and interstitial flow velocities are $\sim1\,\mu\mathrm{m/s}$, the interstitial fluid transport over one heartbeat is small compared with organ-scale transport distances, which justifies a steady-flow model. 
At the inlet and outlet ($\Gamma_i$ and $\Gamma_o$), we prescribe traction boundary conditions to drive the blood flow:
\begin{alignat}{2}
    \boldsymbol{\sigma}_S \bsn &= -p_S^{i}\,\bsn
\quad &&\text{on } \Gamma_i, \\
\boldsymbol{\sigma}_S \bsn &= -p_S^{o}\,\bsn
\quad &&\text{on } \Gamma_o,
\end{alignat}
where $\boldsymbol{\sigma}_S = -p_S \mathbf{I} + \mu_S\big(\nabla \bsu_S + \nabla \bsu_S^{\mathsf{T}}\big)$ is the Cauchy stress tensor for Stokes flow.
We set the inlet pressure $p_S^{i}$ to the mean pressure over the cardiac cycle at the HH15 developmental stage \cite{hu_hemodynamics_1989}. We take $p_S^{o}=0$. 

We model interstitial flow as incompressible porous media flow using Darcy's law \cite{wiig_interstitial_2012}: 
\begin{equation}\label{eq: darcy conservation of mass}
    -\nabla\cdot \left( \frac{k}{\mu_D}\nabla p_D \right) = 0 \quad  \text{in } \Omega_D, 
\end{equation}
Because $\Gamma_D$ comprises anatomical barriers and the symmetry boundary defined in Sec.~\ref{sec: ROI}, we impose zero normal interstitial fluid flux:
\begin{alignat}{1} \label{eq: Zero normal IF flux}
    \bsu_D\cdot\bsn = 0 \quad \text{on } \Gamma_D.
\end{alignat}

At the vessel wall $\gamma$, we impose three interface conditions.
First, we model the filtration of blood plasma into the interstitial region using Eq.~\eqref{eq: Starling's law}; see \cite{wiig_interstitial_2012}.
The spatially uniform oncotic pressure is absorbed into the effective pressure difference. 
Second, we impose the no-slip condition for the blood flow at the vessel wall:
\begin{equation} \label{eq: no slip}
    \bsu_S\cdot \boldsymbol{\tau}_i = 0 \quad \text{ on } \gamma,\quad i=1,2,
\end{equation} 
where $\{\boldsymbol{\tau}_1,\boldsymbol{\tau}_2\}$ are two orthonormal tangent vectors spanning the tangent plane at $\gamma$.
Third, we neglect the filtration of the blood plasma in the blood-flow problem and impose an impermeability condition:
\begin{equation} \label{eq: impermeable blood}
    \bsu_S\cdot\bsn= 0 \quad \text{on } \gamma. 
\end{equation} 
We make this approximation because the filtration velocity is much smaller than the blood velocity.  
In the physiological reference simulation, the peak transvascular flow velocity is $3.0\times 10^{-8}\,\mathrm{m/s}$, whereas the characteristic blood velocity is $10^{-3}\,\mathrm{m/s}$, more than four orders of magnitude higher. 

The flow-model parameter values and sources are listed in Table 1.
The blood-flow parameter values define the fixed blood-flow solution used throughout this work. 
The interstitial-flow parameter values are our best estimates for the yolk sac and define the physiological reference interstitial-flow simulation analyzed in Section~\ref{sec: interstitial flow}. Except for the vessel-wall hydraulic conductivity $L_p$, all parameters were directly taken from the cited sources. 
Following \cite{michel_microvascular_1999}, we estimated $L_p$ by approximating the transvascular leakage across the endothelial junctions as Poiseuille flow between two parallel plates.  
We set the plate gap equal to the endothelial junction width and the plate length equal to the endothelial cell length, using values measured in the chick yolk sac by \cite{ghaffari_blood_2017}.  
We then scaled the junction-level leakage by the vessel-wall porosity to obtain an area-averaged leakage across the entire vessel wall. 
 
\begin{table}[h]
    \centering
\begin{tabular}{|c c c c|}
\hline 
Parameter & Description & Value & Reference\tabularnewline
\hline
$\mu_S$ & Blood viscosity  & $3 \times 10^{-3} \text{ Pa s}$ & \cite{midgett_blood_2015} \tabularnewline

$p^i_S$ & Inlet pressure at vitelline artery & $65.32 \text{ Pa }$ & \cite{hu_hemodynamics_1989} \tabularnewline

$k$ & Interstitial tissue permeability & $5.2 \times 10^{-14} \text{ m}^{2}$ & \cite{ghaffari_blood_2017} \tabularnewline

$\mu_D$ & Interstitial fluid viscosity & $1.2 \times 10^{-3} \text{ Pa s}$ & \cite{swartz_interstitial_2007} \tabularnewline

$L_p$ & Hydraulic conductivity of the vessel wall  & $2.7 \times 10^{-9}\,\mathrm{m}\,\mathrm{Pa}^{-1}\,\mathrm{s}^{-1}$ & \cite{ghaffari_blood_2017} 
\tabularnewline
\hline
\end{tabular}
    \caption{Parameter values used in the simulations.}
    \label{tab:my_label}
\end{table}

\subsection{Numerical method}

Treating the vessel wall as impermeable to blood flow enables a one-way coupling:
We first solve the Stokes flow equations [Eq.~\eqref{eq: Blood flow eqs}] to obtain the blood velocity and pressure, and then solve the Darcy equations [Eq.~\eqref{eq: darcy conservation of mass}] for interstitial pressure.
When we solve the Stokes equations, we treat the interface conditions Eqs.~\eqref{eq: impermeable blood} and \eqref{eq: no slip} on $\gamma$ as boundary conditions.
When we solve the Darcy equations, we treat the filtration condition on $\gamma$ [Eq.~\eqref{eq: Starling's law}] as a Robin boundary condition for $p_D$, since $p_S$ is known from the Stokes flow solution.
We do not require the meshes of the vascular and interstitial domains to be conformal at the interface $\gamma$.
To enforce the filtration boundary condition in the interstitial flow model, we transfer $p_S$ computed on the vascular mesh onto the boundary of the interstitial mesh using nearest-point interpolation.

We use the finite element method (FEM) to discretize the equations. We implement the solver using the open-source FEM package FEniCS \cite{logg_dolfin_2010,alnaes_fenics_2015}. 
Since the Stokes equations represent a saddle point problem, the discrete solution space must satisfy the   Ladyzhenskaya-Babuska-Brezzi (LBB) condition \cite{tezduyar_stabilized_1992}. 
To circumvent this requirement, we stabilize the weak form of the Stokes equations using the Galerkin Least Squares formulation \cite{tezduyar_stabilized_1992}.  
This enables us to employ an equal-order finite element pair for velocity and pressure. Hence, we discretize all discrete solution variables using first-order Lagrange elements. 
We solve the resulting linear Stokes and Darcy systems using MUMPS \cite{amestoy_fully_2001} and GMRES \cite{saad_gmres_1986}, respectively.

By solving the Stokes and Darcy equations sequentially on non-conformal meshes, we can independently choose the mesh resolution in the vascular and interstitial domains.  
We use a coarse mesh for the vascular domain to ensure computations with the direct solver MUMPS are tractable in 3D. 
We use a fine mesh for the interstitial domain to resolve the small interstitial thickness. 
Thus, simplifying the two-way coupled Stokes-Darcy model into a one-way coupled model makes the organ-scale flow computations tractable. 

\section*{Declarations}
\subsection*{Acknowledgments}
We thank Dr.~J.~Matthias Starck for providing the avian embryo yolk-sac histology slides used in this study.

\subsection*{Funding}
This research was supported by the National Science Foundation under award no. 2325419, the Human Frontier Science Program under award no. RGP015/2023, and the Research Foundation–Flanders (FWO) under grant no. G0C8823N. This work used the Bridges-2 system at the Pittsburgh Supercomputing Center through ACCESS allocation MCH220014, supported by National Science Foundation grant nos. 2138259, 2138286, 2138307, 2137603, and 2138296.

\subsection*{Author contributions}

Conceptualization and methodology: A.S., H.G., and E.A.V.J.; experimental data collection: E.A.V.J.; formal analysis: A.S. and H.G.; funding acquisition: H.G.; investigation and software: A.S.; supervision: H.G. and E.A.V.J.; visualization: A.S. and D.P.; writing (original draft): A.S.; writing (review and editing): A.S., H.G., E.A.V.J. and D.P. 

\subsection*{Competing interests}
The authors declare that they have no competing interests.

\subsection*{Data and materials availability}
All data needed to evaluate the conclusions in the paper are present in the paper and/or the Supplementary
Materials. The imaging data and computational code supporting the findings of this study are available from the corresponding author upon reasonable request.

\bibliography{references,references-2}

\clearpage
\thispagestyle{empty}

\begin{center}
{\Large\bfseries Supplementary Materials for\par}

\vspace{0.75em}

{\large\bfseries
Interstitial flow in the chick yolk sac exhibits organ-scale patterns driven by segregated leakage and drainage
\par}

\vspace{1em}

Adithya Srinivasan et al.

\vspace{0.5em}

*Corresponding author. Email: hectorgomez@purdue.edu
\end{center}

\vspace{2em}

\noindent
\textbf{This PDF file includes:}

\begin{itemize}
    \item Supplementary Text
    \item Figs. S1 to S4
\end{itemize}

\clearpage

\setcounter{section}{0}
\setcounter{subsection}{0}
\setcounter{figure}{0}
\setcounter{table}{0}
\setcounter{equation}{0}

\renewcommand{\thesection}{S\arabic{section}}
\renewcommand{\thesubsection}{\thesection.\arabic{subsection}}
\renewcommand{\thefigure}{S\arabic{figure}}
\renewcommand{\thetable}{S\arabic{table}}
\renewcommand{\theequation}{S\arabic{equation}}

\renewcommand{\theHsection}{supp.\arabic{section}}
\renewcommand{\theHsubsection}
    {supp.\arabic{section}.\arabic{subsection}}
\renewcommand{\theHfigure}{supp.\arabic{figure}}
\renewcommand{\theHtable}{supp.\arabic{table}}
\renewcommand{\theHequation}{supp.\arabic{equation}}

\section*{Supplementary Text}

\section{Definition of metrics} \label{app: metrics-defn}
In this supplement, we provide formal definitions of the metrics used in the main manuscript.

\subsection{Flow dimensionality metric}
We define the flow dimensionality metric $K_{DV}$ as the fraction of kinetic energy associated with the dorsal--ventral velocity component. We use this metric for blood flow and interstitial flow. The definitions are:
\begin{equation}
K_{DV}
=
\frac{
\int_{\Omega_S} |u_{S,z}|^2\,dV
}{
\int_{\Omega_S} \|\bsu_S\|^2\,dV
},
\end{equation}
and
\begin{equation}
K_{DV}
=
\frac{
\int_{\Omega_D} |u_{D,z}|^2\,dV
}{
\int_{\Omega_D} \|\bsu_D\|^2\,dV
},
\end{equation}
where $u_{S,z}$ and $u_{D,z}$ are, respectively, the dorsal-ventral component of blood velocity and interstitial flow velocity. A large value of $K_{DV}$ indicates that the flow has a substantial 3D component, while small values indicate that the flow is quasi-2D.

\subsection{Interstitial transport length}

To measure the extent of interstitial fluid transport, we compute the lengths of the streamlines of interstitial flow.
To generate these streamlines, we select 10000 random points on the surface mesh of $\gamma$ and compute streamlines using ParaView's streamline filter \cite{ahrens_paraview_2005}.
Since interstitial flow is much faster than transvascular flow, the interstitial velocity at the vessel wall is nearly tangential to the wall.
As a result, some generated streamlines may terminate incorrectly near the vessel wall.  
Therefore, we retain only streamlines that start at a leakage site ($u_T>0$) and terminate at a drainage site ($u_T<0$).
For each retained streamline, we compute its arc length. We define the interstitial transport length $L_T$ as the mean arc length of the longest 25\% of the retained streamlines. To verify that the long-distance transport was not overestimated due to streamline tortuosity, we also computed the end-to-end displacement of each retained streamline. We define $D_T$ as the mean end-to-end displacement of the 25\% of retained streamlines with the largest displacements.
The end-to-end displacement showed the same trends as the streamline arc length.  
Specifically, for the physiological reference simulation, we find $D_T$ equal to 84\% of $L_T$.
Thus, the value of $L_T$ reflects long-range interstitial fluid transport rather than tortuosity.

\subsection{Intervascular length}

We define the intervascular distance $\ell_{\rm iv}$ as a characteristic measure of the spacing between neighboring vessel segments.   
In the vascular centerline plane, the vessel network splits the interstitial domain into disconnected interstitial regions bounded by neighbouring vessel segments. 
We use the size of the interstitial regions as a proxy for the local intervascular spacing. 
We compute the area-equivalent diameter of each region, defined as the diameter of a circle with the same area, and define $\ell_{\rm iv}$ as the mean of these diameters across all intervascular regions. 

\subsection{Average absolute transvascular pressure drop}
 We define the average absolute transvascular pressure drop over the vessel wall as
\begin{equation} 
\langle |\Delta p|\rangle 
= 
\frac{1}{|\gamma|} 
\int_{\gamma} 
|p_S-p_D|\,dA, 
\end{equation} 
where $|\gamma|$ is the area of the vessel wall. 

\subsection{Ratio of interstitial to transvascular flow}
We define the ratio of vessel-wall averaged interstitial fluid speed to vessel-wall averaged transvascular velocity as
\begin{equation}
R_v
=
\frac{
\int_{\gamma} \|\bsu_D\|\,dA
}{
\int_{\gamma} |u_T|\,dA
}.
\end{equation}
We compute the average over the vessel wall, since $u_T$ is only defined on the vessel wall.

\subsection{Definition of the arterio-venous coordinate}
The arterio-venous coordinate $\psi$ labels the interstitial domain relative to the arterial and venous sides of the vascular network. 
Because the yolk sac vasculature forms an immature capillary plexus at HH15, vessel caliber alone does not clearly distinguish arteries, veins, and capillaries \cite{le_noble_flow_2004}.  
Hence, we use the blood pressure $p_S$ to distinguish arteries and veins and define the arterio-venous coordinate. 
We first prescribe $\psi$ on the vessel wall $\gamma$ as
\begin{equation} 
\label{eq:psi_eq} 
\psi(\bsx) =
\frac{p_S^{i} - p_S(\bsx)}{p_S^{i} - p_S^{o}}
\quad \text{on } \gamma.
\end{equation}  
By construction, $\psi = 0$  near the inlet (vitelline artery) and $\psi=1$ near the outlet (vitelline vein). Next, we extend $\psi$ from the vessel wall $\gamma$ into the interstitial domain $\Omega_D$ by solving Laplace's equation
\begin{alignat}{1}  
\Delta\psi = 0 \quad \text{in } \Omega_D,
\end{alignat}  
with a Dirichlet boundary condition on $\gamma$ prescribed by Eq.~\eqref{eq:psi_eq}. 
We impose $\nabla\psi\cdot\bsn=0$ on $\Gamma_D$.
This homogeneous Neumann condition does not prescribe $\psi$ or introduce a source or sink. The Laplace extension is driven solely by the prescribed values of $\psi$ on the vessel wall $\gamma$ and yields a smooth interpolation of $\psi$ from the vessel wall $\gamma$ into the interstitial domain.
By the maximum principle for harmonic functions, this construction avoids introducing artificial extrema. Since $\psi\in[0,1]$ on $\gamma$, we have $\psi\in[0,1]$ throughout $\Omega_D$. 

\subsection{Interstitial fluid flux through arterio-venous isosurfaces}
To quantify the accumulation of the interstitial fluid flux, we compute the interstitial fluid flux through isosurfaces of the arterio-venous coordinate $\psi$.
For a scalar value $\psi_c$, we denote the corresponding isosurface of $\psi$  by
\begin{equation}
\Pi_c
=
\left\{
\bsx\in\Omega_D:\psi(\bsx)=\psi_c
\right\}.
\end{equation}
We define the interstitial fluid flux through $\Pi_c$ as
\begin{equation}
Q(\psi_c)
=
\int_{\Pi_c}
\bsu_D\cdot\bsn_\psi\,dA,
\end{equation}
where  $\bsn_\psi$ is the unit normal to $\Pi_c$, pointing along the direction of increasing $\psi$.
Thus, $Q(\psi_c)$ measures the net interstitial fluid flux crossing the isosurface of $\psi$  from the arterial side toward the venous side.

\section{Vascular information length} \label{app: section_scale}

\begin{figure}
    \centering
    \includegraphics[width=\linewidth]{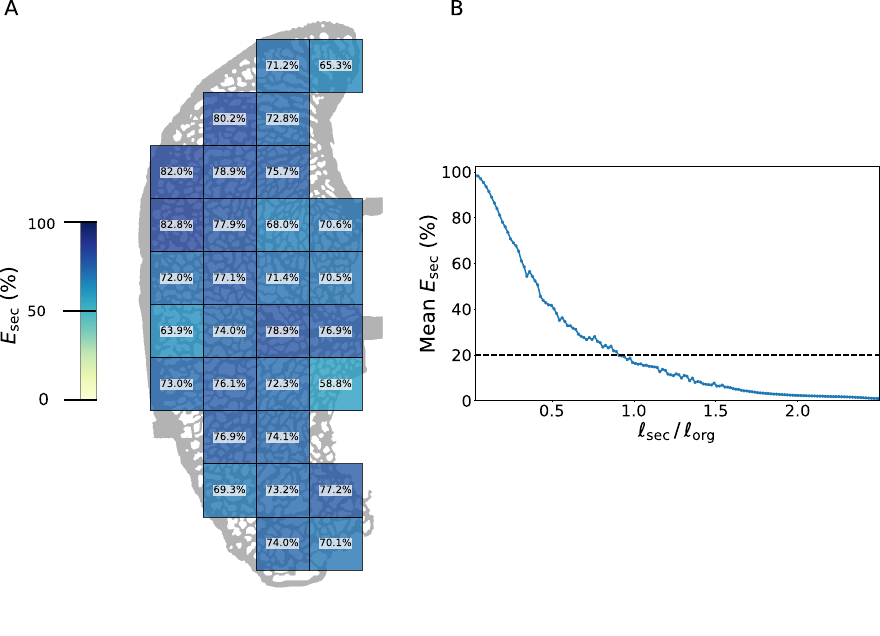}
    \caption{\textbf{Section-scale simulations determine the vascular information length.}
    (A) Normalized section-scale error $E_{\rm sec}$ for the physiological reference simulation. Each tile represents a square organ section overlaid on the binary image of the yolk sac to indicate the section location.  
The tile color and overlaid text indicate $E_{\rm sec}$.  
(B) Plot of the mean $E_{\rm sec}$ against the section size $\ell_{\rm sec}$ normalized by the yolk-sac extent along the distal-proximal axis $\ell_{\rm org}$. The dashed line denotes the $20\%$ error threshold below which the section-scale solution is considered to approximate the organ-scale solution.}
    \label{fig: vascular information length}
\end{figure}

The vascular information length quantifies the amount of vascular context that is required to estimate the organ-scale solution. We use section-scale simulations to determine the vascular information length. We define a section-scale domain by selecting a square window with side length $\ell_{\rm sec}$ in the vascular centerline plane and extruding it in the dorsal-ventral direction. 
We generate multiple section-scale domains by varying the location and size of this window.   
The interstitial portion of a section-scale domain is denoted by $\Omega_D^{\rm sec}$. 
The boundary of $\Omega_D^{\rm sec}$ is the union of the vessel wall inside the section, $\gamma^{\rm sec}$, the outer interstitial boundary inside the section, $\Gamma_D^{\rm sec}$, and the artificial cut boundary introduced by the rectangular section, $\Gamma_{\rm cut}$. In the section-scale studies, we use the blood pressure computed from the organ-scale simulation to drive the interstitial flow. 
We solve the same interstitial-flow model as in the organ-scale simulation, but the interstitial domain is restricted to $\Omega_D^{\rm sec}$.  
We impose the filtration law [Eq.~\eqref{eq: Starling's law}] on $\gamma^{\rm sec}$ and no-flux boundary conditions on $\Gamma_D^{\rm sec}\cup\Gamma_{\rm cut}$. The no-flux condition on $\Gamma_{\rm cut}$ ensures that interstitial flow outside the simulated tissue section does not influence the solution inside the section.  
We denote the interstitial fluid velocity obtained from these simulations by $\bsu_D^{\rm sec}$. 
To avoid remeshing the geometry for each section-scale domain, we perform the section-scale simulations using the diffuse-domain method \cite{stein_computational_2017, srinivasan_computational_2026}. 
To compare the section-scale solution and the organ-scale solution, we compute the normalized error in interstitial fluid velocity:
\begin{equation}
E_{\rm sec}
=
\frac{
\int_{\Omega_D^{\rm sec}}
\left\|
\bsu_D^{\rm sec}-\bsu_D
\right\|\,dV
}{
\int_{\Omega_D^{\rm sec}}
\left\|
\bsu_D
\right\|\,dV
}.
\end{equation}
Since $E_{\rm sec}$ depends on the location of the section, we perform section-scale simulations at multiple locations for each section size. 
Fig.~\ref{fig: vascular information length}A shows $E_{\rm sec}$ for the physiological reference simulation at different locations for the section size corresponding to the representative section shown in the inset of Fig.~\ref{fig:Interstitial flow}. 
We summarize the error for the section size using the mean error across all locations. 
For this section size, the mean error across the sampled locations is 73.5\%.

We next examine how $E_{\rm sec}$ varies with section size by varying $\ell_{\rm sec}$ over $[\ell_{\rm iv},2.5\;\ell_{\rm org}]$ in increments of $0.5 \; \ell_{\rm iv}$, where the upper bound corresponds to the anterior-posterior extent of the yolk sac. 
We define eligible windows as those whose overlap with the combined vascular and interstitial domain exceeds $75\%$. 
We use the following strategy to select window locations for each section size: 
First, we select the maximum number of non-overlapping eligible windows. 
If fewer than ten non-overlapping windows are available, we allow overlap and select ten eligible windows whose union provides the greatest spatial coverage of the domain. 
This strategy provides at least ten windows for each section size while maximizing spatial coverage, thereby reducing bias associated with both the number and spatial distribution of the sampled windows. 
When the window size $\ell_{\rm sec} > L_{\rm DP}$, where $L_{\rm DP}$ is the distal-proximal length, a square window cannot be placed within the domain. For these section sizes, we therefore use rectangular windows with side lengths $\ell_{\rm sec}$ and $L_{\rm DP}$. 
Fig.~\ref{fig: vascular information length}B shows that the mean $E_{\rm sec}$ decreases with increasing size because a larger section contains more vascular information to better reproduce the interstitial flow. 
The trend is not strictly monotonic because the sampled window locations change between section sizes. 
We consider a section-scale simulation to reproduce the organ-scale solution when the mean $E_{\rm sec}<20\%$.
Using this criterion, we define the vascular information length $L_I$ as the smallest section size for which the mean $E_{\rm sec}$ falls below $20\%$.  
From Fig.~\ref{fig: vascular information length}B, we obtain $L_I = 0.91 \; \ell_{\rm org}$ for the physiological reference simulation. 
We use the same procedure to compute $L_I$ in the sensitivity analysis described in Section~\ref{sec : sensitivty analysis}.

\begin{figure}
    \centering
    \includegraphics[width=\linewidth]{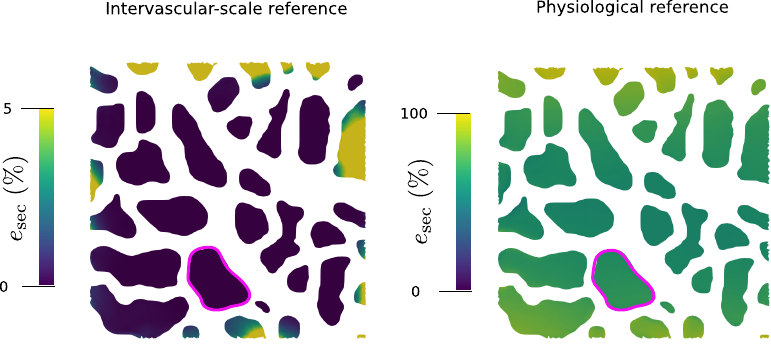}
    \caption{
    Local normalized velocity difference between the organ-scale and section-scale simulations $e_{\rm sec}(\bsx)$  for the intervascular-scale reference and the physiological reference, plotted on a 2D section in the vascular centerline plane. 
In this plane, interstitial regions are disconnected. The magenta boundary marks one such interstitial region that is isolated from the boundary.}
    \label{fig: excess-info}
\end{figure}

To identify where the section-scale approximation fails, we compute the local normalized velocity difference between the organ-scale and section-scale simulations as 
\begin{equation} 
e_{\rm sec}(\bsx) 
= 
\frac{ 
\left\| 
\bsu_D^{\rm sec}(\bsx)-\bsu_D(\bsx) 
\right\| 
}{ 
\|\bsu_D\| 
}. 
\end{equation} 
Fig.~\ref{fig: excess-info} shows $e_{\rm sec}(\bsx)$ for the intervascular-scale reference and the physiological reference simulations at the representative section-scale domain shown in the inset of Fig.~\ref{fig:Interstitial flow}. 
In the intervascular-scale reference, $e_{\rm sec}$ is nearly zero in the isolated interstitial regions that do not intersect the artificial cut boundary and is concentrated primarily in regions cut by $\Gamma_{\rm cut}$. 
This observation is consistent with intervascular-scale flow patterns, where streamlines remain confined to individual interstitial regions and are determined mainly by the nearby vascular network. 
In contrast, the physiological reference simulation shows elevated $e_{\rm sec}$ even in isolated interstitial regions away from the artificial cut boundary.  
This indicates that flow in these regions is influenced not only by nearby vessels, but also by leakage and drainage elsewhere in the yolk-sac network.

\section{Sensitivity analysis based on the conductivity ratio $R_c$} \label{app: sensitivty_analysis}
In this section, we describe the details of the sensitivity analysis performed in Sec.~\ref{sec : sensitivty analysis}. 
For a fixed $p_S$, the interstitial flow solution is determined by  Eqs.~\eqref{eq: darcy conservation of mass}, \eqref{eq: Starling's law}, and \eqref{eq: Zero normal IF flux}.  
We rewrite these equations as
\begin{alignat}{3} 
-\nabla\cdot\left(\nabla p_D\right) &= 0 
\quad &&\text{in } \Omega_D, \\ 
-R_c \nabla p_D \cdot \bsn &= \frac{p_S - p_D}{t_w} 
\quad &&\text{on } \gamma, \label{eq: Rp interface}\\ 
\nabla p_D\cdot\bsn &= 0 
\quad &&\text{on } \Gamma_D.
\end{alignat}
In this form, the interstitial flow problem depends on the hydraulic parameters only through $R_c$.  
Therefore, for a fixed $p_S$, the interstitial pressure $p_D$, the interstitial transport length $L_T$, and the vascular information length $L_I$ depend on the interstitial flow parameters $(L_p,k,\mu_D)$ only through $R_c$.
Thus, varying $R_c$ in the sensitivity analysis captures the variation of all the interstitial-flow parameters.
We vary $R_c$ over the range $[0.001R^{\rm phy}_c, 1000 R^{\rm phy}_c]$, where  $R^{\rm phy}_c$ is the value of $R_c$ in the physiological reference simulation. 
This range was chosen to include both the intervascular-scale reference simulation and the physiologically plausible range of flow parameters for the chick yolk sac. 
In this range, we perform simulations at 40 values of $R_c$ that are equally spaced on a logarithmic scale.

To determine the physiologically plausible range for $R_c$, we use bounds on the vessel-wall hydraulic conductivity $L_p$, the interstitial permeability $k$, and the interstitial fluid viscosity $\mu_D$. 
In our estimate for $L_p$, the junction width $d_j$ is the most critical parameter.   
In our physiological reference simulation, we compute $L_p$ with $d_j =70$ nm, as reported in \cite{ghaffari_blood_2017}.   
In \cite{ghaffari_blood_2017}, $d_j$ is measured at a slightly earlier developmental stage than in our simulation (HH10 vs HH15).  
As the embryo develops, the endothelial junctions mature, and junction width generally decreases.  
To account for this possibility, we estimate the lower bound of $L_p$ using $d_j = 20$ nm, which is commonly reported for mature vasculature \cite{michel_microvascular_1999}.  
To estimate an upper bound, we use $d_j = 80$ nm as a plausible exploratory value representing vasculature that is more permeable than in the physiological reference simulation.  
Such increased leakiness can arise, for example, when the glycocalyx is degraded \cite{henderson-toth_glycocalyx_2012}.

The permeability used in the physiological reference simulation, denoted by $k^{\rm phy}$, is taken from \cite{ghaffari_blood_2017}.  
However, unlike $L_p$, we cannot systematically relate variation in permeability to an underlying measured geometric parameter.  
Therefore, we consider the physiological range as $[0.1\,k^{\rm phy},\;10\,k^{\rm phy}]$.  
This range agrees with the interstitial permeability values reported for collagen gels \cite{cacheux_asymmetry_2023}, which are widely used as ECM-mimicking materials.  
We keep $\mu_D$ fixed because its physiological variation is expected to be small compared with the uncertainty in $L_p$ and $k$.  
From the bounds on $k$ and $L_p$, we determine the physiological range of $R_c$ as $  \left[  
\frac{\min(k)}{\mu_D \max(L_p)\, t_w},  
\;  
\frac{\max(k)}{\mu_D \min(L_p)\, t_w}  
\right]  
$.  

\section{Variation of flow pattern metrics with the conductivity ratio $R_c$}
\label{app: Rc_metrics}
\begin{figure}
    \centering
    \includegraphics[width=\linewidth]{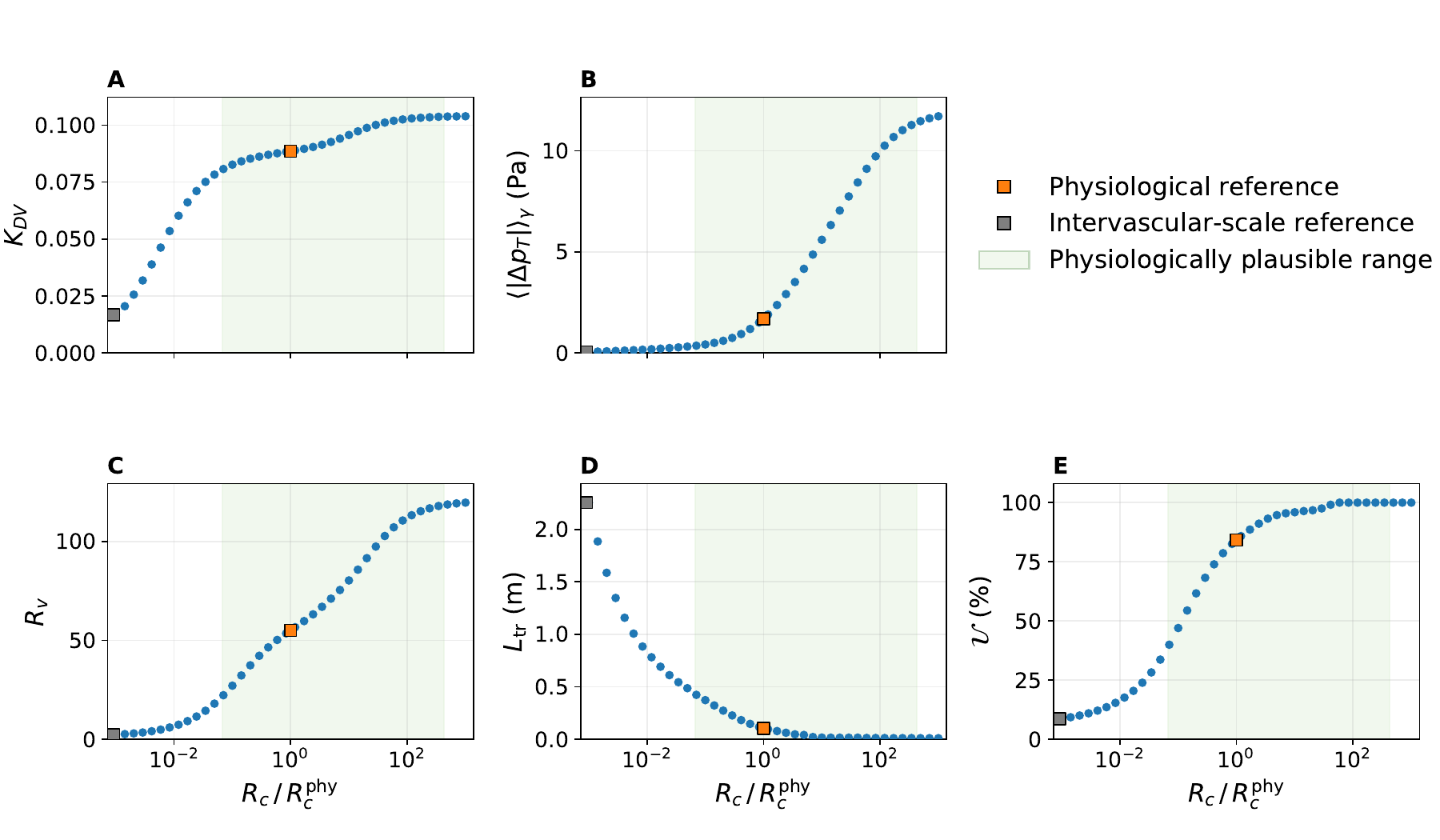}
    \caption{\textbf{Variation of interstitial flow-pattern metrics with the conductivity ratio $R_c$.} (A) Flow dimensionality metric $K_{DV}$. (B) Average transvascular pressure drop $\langle |\Delta p|\rangle$. (C) Ratio of interstitial to transvascular velocity $R_v$. (D) Length of the $u_T=0$ contour, $L_{\rm tr}$. (E) Utilization of transvascular leakage $\mathcal{U}$. 
    }
    \label{fig: R_c metrics}
\end{figure}

In this section, we examine how the signatures of interstitial flow vary with the conductivity ratio $R_c$.
Fig.~\ref{fig: R_c metrics}A--C shows that the flow dimensionality metric $K_{DV}$, the average transvascular pressure drop $\langle|\Delta p|\rangle$, and the ratio of average interstitial flow speed to transvascular speed increase monotonically with $R_c$. 
These trends support the interpretation of all three metrics as signatures of the spatial scale of interstitial flow. 

Next, we quantify the strength of spatial segregation of leakage and drainage and the accumulation of the interstitial flux. 
We quantify segregation between leakage and drainage using $L_{\rm tr}$, the length of the $u_T=0$ contour on the vessel wall $\gamma$. 
A smaller $L_{\rm tr}$ indicates that the transition from leakage to drainage occurs over a smaller portion of $\gamma$, and therefore corresponds to stronger spatial segregation. 
In the physiological reference simulation, $L_{\rm tr}=0.1~\mathrm{m}$, compared with $L_{\rm tr}=2.25~\mathrm{m}$ in the intervascular-scale reference, indicating stronger segregation between leakage and drainage in the physiological reference case. 
To quantify interstitial flux accumulation, we define the utilization $\mathcal{U}$ as
\begin{equation}
\mathcal{U}
=
100
\frac{
\max_{\psi_c} Q(\psi_c)
}{
\int_{\gamma_+} u_T\,dA
},
\end{equation}
where
\begin{equation}
\gamma_+
=
\left\{
\bsx\in\gamma:\,u_T(\bsx)>0
\right\}
\end{equation}
is the leakage region of the vessel wall. 
Thus, $\mathcal{U}$ is the maximum percentage of the total leaked flux that passes through any isosurface of $\psi$. 
The limiting case $\mathcal{U}=100\%$ occurs when all leaked interstitial fluid flux passes through a single isosurface of $\psi$ before being drained. 
We find $\mathcal{U}=84\%$ for the physiological reference simulation, compared with $\mathcal{U}=8.7\%$ for the intervascular-scale reference, indicating a stronger accumulation of interstitial flux in the physiological reference simulation. Consistent with this stronger accumulation, the ratio of interstitial fluid speed to transvascular velocity $R_v$ is substantially higher in the physiological reference case than in the intervascular-scale reference. 
Fig.~\ref{fig: R_c metrics}D and E show that $L_{\rm tr}$ decreases while $\mathcal{U}$ increases with the conductivity ratio. 
Together, these trends indicate that the transition toward organ-scale flow organization is accompanied by stronger segregation of the transvascular filtration and greater interstitial flux accumulation.

\section{Interstitial-flux accumulation amplifies interstitial fluid speed along a representative flow path}
\label{app: flux accumulation}

\begin{figure}
    \centering
    \includegraphics[width=\linewidth]{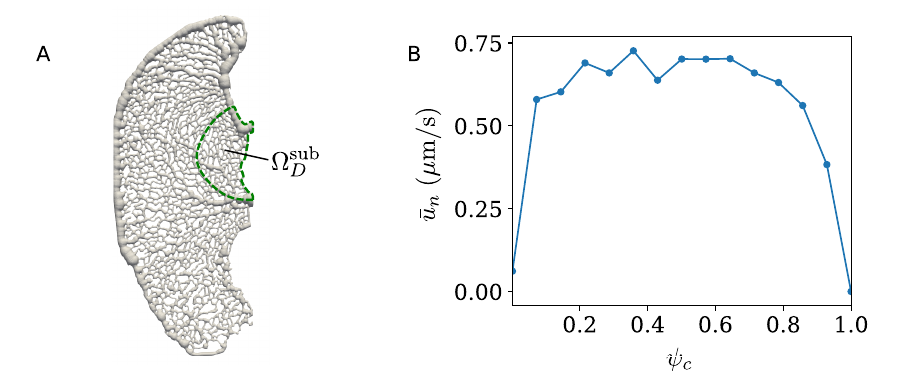}
    \caption{\textbf{Interstitial flux accumulation amplifies flow speed along an artery-to-vein pathway.}
    (A) The subregion $\Omega^{\rm sub}_D$ used for this analysis is indicated on the vascular network.  The green dashed line marks the boundary of $\Omega^{\rm sub}_D$ in the vascular centerline plane. We extrude this boundary through the tissue thickness to define the 3D domain of $\Omega^{\rm sub}_D$. 
(B) Area-averaged normal component of interstitial fluid velocity $\bar{u}_n$ across the restricted isosurface of the arterio-venous coordinate, given by $\psi=\psi_c$. The isosurfaces are restricted to $\Omega^{\rm sub}_D$.
    }
    \label{fig: subregion1_IF_TF}
\end{figure}

To directly connect the accumulation of interstitial flux with interstitial fluid speed, we analyze the average interstitial fluid speed along a representative flow path.
Interstitial flow has two dominant flow paths with different interstitial fluid speeds.
Computing the average normal velocity on the full isosurfaces of $\psi$ would mix velocities from both paths.
To avoid mixing these paths, we restrict the analysis to a subregion of the yolk-sac tissue, $\Omega_D^{\rm sub}$, that contains the interstitial flow path from the vitelline artery to the vitelline vein (Fig.~\ref{fig: subregion1_IF_TF}A).
We choose this subregion because interstitial flow and transvascular flow attain their highest speeds in this region.  

Let $\Pi_c^{\rm sub}=\Pi_c\cap\Omega_D^{\rm sub}$ denote the isosurface of $\psi$ restricted to $\Omega_D^{\rm sub}$. 
We compute the area-averaged normal component of the interstitial fluid velocity over the restricted isosurface $\Pi_c^{\rm sub}$ as
\begin{equation}
\bar{u}_n(\psi_c)
=
\frac{1}{|\Pi_c^{\rm sub}|}
\int_{\Pi_c^{\rm sub}}
\bsu_D\cdot\bsn_{\psi}\, dA,
\end{equation}
where $|\Pi_c^{\rm sub}|$ is the area of $\Pi_c^{\rm sub}$, and $\bsn_{\psi}$ is the unit normal to $\Pi_c$, pointing in the direction of increasing $\psi$.

Fig.~\ref{fig: subregion1_IF_TF}B plots $\bar{u}_n(\psi_c)$ against $\psi_c$. 
At the start of the flow path ($\psi=0.0016$), the average normal velocity is $0.06~\mu\text{m}/\text{s}$, which is comparable to the local transvascular leakage velocity. 
As we move along the arterio-venous coordinate, $\bar{u}_n$ rises to $\approx 0.70~\mu$m/s at $\psi =0.5$, representing an 11-fold increase.
This increase in interstitial fluid speed follows the same trend as the interstitial flux $Q$.
This correspondence suggests that interstitial fluid flux accumulation amplifies the interstitial fluid speed along this representative path, allowing interstitial flow to exceed the local transvascular flow.

\end{document}